\documentclass[manuscript]{aastex}

\usepackage{apjfonts}
\usepackage{emulateapj5}
\usepackage{amsmath}
\usepackage{epsfig}
\usepackage{amssymb}

\renewenvironment{figure}{\begin{figure*} }{\end{figure*}}

\newcommand{\rhos}{\rho_{\rm s}}
\newcommand{\rhop}{\rho_{\rm p}}

\newcommand{\dd}{{\rm d}}

\newcommand{\smax}{s_{\rm max}}
\newcommand{\smin}{s_{\rm min}}
\newcommand{\nmax}{n_{\rm max}}
\newcommand{\mmax}{m_{\rm max}}
\newcommand{\Stmax}{St_{\rm max}}
\newcommand{\hp}{h_{\rm p}}
\newcommand{\Sigmap}{\Sigma_{\rm p}}
\newcommand{\alphat}{\alpha_{\rm t}}

\newcommand{\ve}{v_{\rm e}}
\newcommand{\taud}{\tau_{\rm d}}
\newcommand{\rau}{r_{\rm AU}}
\newcommand{\up}{u_{\rm p}}
\newcommand{\Dp}{D_{\rm p}}
\newcommand{\nut}{\nu_{\rm t}}
\newcommand{\hbarau}{\overline{h}_{\rm AU}}
\newcommand{\Gp}{G_{\rm p}}

\newcommand{\Sceff}{Sc_{\rm eff}}
\newcommand{\rhill}{r_{\rm H}}

\begin{document}

\title{Growth and migration of solids in evolving protostellar disks I: Methods \& Analytical tests}
\author{P. Garaud} 

\affil{Department of Applied Mathematics and Statistics, Baskin School of Engineering, University of California
Santa Cruz, 1156 High Street, CA-95064 Santa Cruz, USA} 

\maketitle

\begin{abstract}
This series of papers investigates the early stages of planet formation by 
modeling the evolution of the gas and solid content of protostellar disks 
from the early T Tauri phase until complete dispersal of the gas. 
In this first paper, I present a new set of simplified equations modeling the 
growth and migration of various species of grains in a gaseous protostellar 
disk evolving as a result of the combined effects of viscous accretion 
and photo-evaporation from the central star.
Using the assumption that the grain size distribution function always 
maintains a power-law structure approximating the average outcome of the 
exact coagulation/shattering equation, the model focuses on the calculation 
of the growth rate of the largest grains only. The coupled evolution 
equations for the maximum grain size, the surface density of the gas 
and the surface density of solids are then presented and solved 
self-consistently using a standard 1+1 dimensional formalism. 
I show that the global evolution of solids is controlled 
by a leaky reservoir of small grains at large radii, and
propose an empirically derived evolution equation for the total mass 
of solids, which can be used to estimate the total heavy element retention 
efficiency in the planet formation paradigm. Consistency with observation 
of the total mass of solids in the Minimum Solar Nebula augmented 
with the mass of the Oort cloud sets strong upper limit on the 
initial grain size distribution, as well as on the turbulent parameter 
$\alphat$. Detailed 
comparisons with SED observations are presented in a following paper. 

\end{abstract}


\section{Introduction}

\subsection{Theoretical and observational motivations}

More than two hundred extrasolar planets have now been detected, revealing 
surprising diversity. Doppler surveys have
provided a large database of masses, orbital radii and
eccentricities,  which show notably few (and a few notable)
systematics, as for example  the relationship  between stellar
metallicity and the number of detected planets (Fischer \& Valenti,
2005).  Transit detections are now also beginning to show a large
diversity in the  internal structure of planets with otherwise very similar
properties (Guillot {\it et al.}, 2006).

Fast-forwarding back a few Gyr, one can rightfully expect to find the 
origin of exo-planetary diversity in the equivalent diversity of 
protostellar disks. And evidence has indeed been found to support this idea.
The observed fraction of stars showing excess at near-IR 
(Haisch {\it et al.} 2001, Hartmann {\it et al.} 2005, 
Sicilia-Aguilar {\it et al.} 2006) and/or mid-IR 
wavelengths (Mamajek {\it et al.} 2004) steadily decreases from nearly 
100\% for stars within the youngest clusters, to zero for stars within 
clusters older than about 20 Myr. This correlation has long been 
interpreted as clear evidence for disk dispersal within a typical
timescale of about 10Myr, but is now beginning to gather additional interest 
as evidence for a large variation in the disk dispersal rates amongst 
similar type stars within the same cluster. This dispersion could be 
related to variations in the initial disk conditions 
and/or to the characteristics of the host star 
(Hueso \& Guillot, 2005). Other possible tracers of disk structure and/or 
evolution (such as the crystallinity fraction and grain growth) also 
reveal significant diversity: for instance, co-eval stars of similar types 
show evidence for very different 
crystallinity fractions (Meeus {\it et al.} 2003 for T Tauri stars, 
Apai {\it et al.} 2005 for brown dwarves). 

Can the origin of this dynamical and structural diversity indeed be
traced back to the initial conditions of the disk? 
Qualitatively speaking, 
can it explain why some systems form planets while others 
don't? Quantitatively speaking, is there
a link between the initial angular momentum and mass of the disk and the 
characteristics of the emerging planetary system?

Meanwhile, stringent upper bounds
on the total amount of heavy elements typically remaining as planetary building
blocks have been deduced from the very low metallicity dispersion 
measured amongst similar type stars within the same cluster by
Wilden {\it et al.} (2002). This result is puzzling in the light 
of the contrastingly large range of observed disk
survival timescales: how can widely different
dynamics lead to similar retention efficiencies.

A necessary step towards answering these questions 
is the development of a comprehensive numerical model capable of 
following the formation and evolution of planetary systems from their 
earliest stages to the present day, including all of the physical 
processes currently  understood to play a role in the evolution of 
the gas and solids.

The standard core-accretion model of planet formation begins with the
condensation of heavy elements into small grains, 
followed by their stochastic collisional growth into 
successively larger aggregates until they reach a typical mass 
(either collectively or individually) where mutually induced gravitational 
forces begin to influence their motions. The small planetesimals then 
continue growing by accreting each other (together with some of the disk gas),
 until a critical point is reached where runaway gas accretion may eventually 
begin. This first planetary formation phase ends with the dispersal of the 
disk gas, possibly by photo-evaporation, although gravitational interactions 
between the various bodies continue taking place resulting in close 
encounters (sometimes collisions) with dynamical rearrangement of the system 
(including ejection, shattering, coagulation).
 
In this paper I present a numerical model for the first stage of this
process, in which a protostellar disk and all of its contents (both in 
gaseous and in solid form) are evolved simultaneously until complete 
dispersal of the gas. The next stages 
of evolution from this point onward are best treated with an N-body code, for 
which the results presented here could be used as initial conditions.

Recent data obtained with the Spitzer Space Telescope has provided
valuable information on the evolution of grains in protostellar disks, 
which can be used to both construct and test the desired planet formation 
model. Since the near- and mid-IR ranges of the observed spectral energy 
distributions (SEDs) are essentially due to reprocessing 
of the stellar radiation by small dust grains, the key to modeling 
planet formation in the context of evolving disks is to better understand 
the relationship between the observable SEDs and the physics which couple 
the gas and dust dynamics under the gravitational and radiative influence 
of a central star. This is done in Paper II (Alexander \& Garaud, 2007).
 
\subsection{General methodology}

This work presents a new versatile numerical tool to 
study the evolution of both gas and solids in protostellar disks, 
from classical T Tauri disks to transition disks 
and finally to forming planetary systems (embedded perhaps in a debris disk).  
The model developed takes into account the following physical 
phenomena: (i) axisymmetric 1+1D gas dynamics around the central star, 
(ii) photo-evaporation by the central star,
(iii) continuous grain size distribution maintained by growth 
and fragmentation, 
(iv) grain sublimation and condensation, 
(v) multiple grain species (iron, silicates, ices), 
(vi) gas-grain coupling including turbulent dust suspension, 
turbulent diffusion and drift and
(vii) gravitational interaction between forming embryos 
(in a statistical sense). 

While the general goal of modeling the early disk evolution has been 
pursued by many others before, this particular model is the 
first to include all of the physics listed above in a single, 
well-tested, fast and practical algorithm. Other physical phenomena such as 
photo-evaporation by nearby stars, truncation of the disk by stellar fly-by, 
or planetary migration are easy to implement, but not discussed here. 
In order to place the model in context, it is useful to summarize 
briefly existing work on the subject. A more thorough discussion of 
the results in the light of previous work can be found in 
\S\ref{sec:discuss}.

Axisymmetric {\it gas} dynamics in a viscously dominated accretion disk has 
been thoroughly analyzed by Lynden-Bell \& Pringle (1974). In subsequent
work, particular attention was given to studying the disk structure and 
evolution in the light of SED observations (see Hartmann 
{\it et al.} 1998 for example). Photo-evaporation of 
the gas by UV photons (either ambient and/or emerging from central star)
is now thought to play a major role in the dispersal of the disk gas. This 
was studied in detail by Hollenbach {\it et al.} (1994), and later proposed
 by Clarke, Gendrin \& Sotomayor (2001) as a possible model providing the 
characteristic ``two-timescale'' evolution (namely a long lifetime 
with a rapid dispersal 
time) required by the low relative abundance of transition disks (see the reviews by Hollenbach \& Gorti 2005, and Dullemond {\it et al.} 2007).  

Meanwhile, the study of the evolution of {\it solids} in protostellar disks 
also has a long history, where the particular emphasis has in the 
vast majority of cases been to model the formation of our own solar system.
The early works of Whipple (1972) and Weidenschilling (1977) laid the 
foundation for studying the motion of small solid bodies in the early 
solar nebula. Voelk {\it et al.} (1980) developed a theory for the 
dynamical coupling of solid particles with turbulent eddies, which enabled 
many further studies of the collisional growth of dust grains into 
planetesimals (Weidenschilling, 1984 and subsequent papers, 
Weidenschilling \& Cuzzi 1993, Stepinski \& Valageas 1997, 
Suttner \& Yorke 2001, Dullemond \& Dominik, 2005). Finally, steady 
progress in the interpretation of various cosmochemistry data has 
prompted the need for a better understanding of the
evolution of the various chemical species present in the disk, in particular
water. In addition to their own work, Ciesla \& Cuzzi (2006) present 
an excellent review of recent advances in the field. 

Combining the evolution of solids with the evolution of the gas with 
the aim of bridging the gap between SED interpretations and our own 
solar system formation is naturally the next step in this scientific 
exploration process. The work of Suttner \& Yorke (2001) pioneered the
concept when looking at grain growth and migration in the very early 
stages of the disk formation (first few $10^4$ yr). Alexander \& Armitage 
(2007) (AA07 hereafter) were recently the first to combine state-of-the-art 
photo-evaporation models with grain migration to gain a better 
understanding of the nature of some forming transition disks. 
The proposed model draws from many of the fundamental ideas of 
these previous studies; in particular, it can be thought of as 
a generalization of the AA07 model which includes the effects of grain 
growth, sublimation and condensation. 

Theoretical studies of dust growth typically require the solution of a
collisional equation at every spatial position of the disk. Amongst
some of the difficulties encountered one could mention the
determination of the particle structure, the sticking efficiency, the
shattering threshold and the size distribution of the fragments, 
and not least the relative velocities of the particles before
collision. Indeed, while the motion of particles in a laminar disk is
fairly easy to compute, matters are complicated when dynamical
coupling between grains and turbulent eddies is taken into
account. Tiny grains are well-coupled with the gas though
frictional drag, while larger ``boulders'' only feel the eddies as a random
stochastic forcing. The intrinsic dispersion and the relative
velocities of the particles can be modeled statistically
provided one assumes the gas eddies follow a turbulent Kolmogorov
cascade from the macro-scale to the dissipation scale. This idea was
originally proposed by Voelk {\it et al.} (1980) and more recently reviewed by
various authors, notably Weidenschilling (1984). Yorke \& Suttner (2001)
and Dullemond \& Dominik (2005) used these velocity prescriptions to
evaluate the rate of growth of particles in protostellar disks by
solving the full coagulation equation. Their results show that 
the collisional growth of particles in the inner regions of the disk 
is too fast, unless shattering is taken into account. It is therefore
vital to include it in evolutionary models of disks as well.

However, solving for the complete coagulation/shattering equation
for every particle size, at every timestep and for every position in the 
disk is computationally prohibitive. Statistical surveys of the typical
outcome of the disk evolution for a wide range of stellar parameters
and initial conditions cannot be done in this fashion. 

The novel part of this work concerns the modeling of the 
evolution of the grain size distribution function under collisional 
coagulation and shattering. The underlying
assumption of the model proposed is that collisions between dust grains are
frequent enough for a quasi-steady coagulation/shattering balance to
be achieved in such a way as to maintain a power-law particle size
distribution function with index $-3.5$ as in the ISM, but with
varying upper size cutoff $\smax$. With this assumption, the study of
the evolution of solids in the disk can be reduced to a small 
set of one-dimensional partial differential equations for the maximum
particle size $\smax(r,t)$, the total surface density of gas 
$\Sigma(r,t)$, as 
well as the total surface density of solids and vapor for each 
species considered ($\Sigmap^i(r,t)$ and $\Sigma^i_{\rm v}(r,t)$, where 
$i$ is the index referencing the species). 
Here $r$ is the radial distance from the central 
star and $t$ is time. This idea is to be considered as an alternative 
approach to the work of 
Ciesla \& Cuzzi (2006) for instance, who equivalently model the 
evolution of gas and solids in the disk over the course of several 
Myr, simplifying the collision/shattering balance by considering only 
four ``size'' bins (vapor, grains, rapidly drifting ``migrators'' and 
finally very large planetesimals).

\subsection{Outline of the paper}

The derivation of the model is presented in complete 
detail in \S\ref{sec:model} (the result-minded reader may prefer to jump 
straight to \S\ref{sec:fiducial} and \S\ref{sec:results}).
The standard gas dynamics equations together with the photo-evaporation 
model used are well-known, and summarized for completeness in 
\S\ref{subsec:modelgas} and \S\ref{subsec:modelvap}.
The basic assumptions for the 
particle size distribution model considered as the basis for this paper 
are presented in \S\ref{subsec:sizefunc}. The stochastic motion of solids 
in the nebula resulting from frictional coupling 
with turbulent eddies and from mutual gravitational encounter have 
been studied by many others before. Key results from these works are
presented in \S\ref{subsec:motion}, and later used in \S\ref{subsec:growth}
and \S\ref{subsec:sigmapevol} to derive new equations for the growth of 
grains into planetesimals, as well as the evolution of the total 
surface density of particles. Finally, \S\ref{subsec:sublim} 
summarizes the very simple sublimation/condensation model used here.

A general overview of the typical inputs and outputs of the numerical model 
are given in \S\ref{sec:fiducial} and \S\ref{sec:results} 
respectively. In order to gain a better understanding 
of the numerical results, \S5 presents existing and new analytical work 
characterizing the global features of the model (gas dynamics in 
\S\ref{subsec:mathgas}, grain growth in \S\ref{subsec:mathgrow}, 
evolution of solids in \S\ref{subsec:mathsolids}, \S5.4 and \S5.5). 
In particular, a plausible new semi-analytical evolution equation for the 
total mass of solids in the disk is presented in \S\ref{subsubsec:totalmassp}, 
which depends only on the initial conditions of the disk. Finally, the 
model and results are discussed in \S\ref{sec:discuss}. Although this paper 
focuses primarily on presenting the methods used (while paper II discusses the 
observable properties of the modeled disks), I give some estimates for 
the heavy-element retention efficiency of disks as a function of the 
model parameters, and show how one could reconcile the high diversity of 
observed disk properties with the low dispersion in metallicities for
star within the same cluster (Wilden {\it et al.} 2002). Conclusions 
are summarized in \S\ref{sec:conclude}.

\section{Model setup}
\label{sec:model}

\subsection{Evolution of the gas disk}
\label{subsec:modelgas}

In all that follows, I assume that the gas disk evolves independently of 
the solids. Note that this is only true as long as the surface density 
of the gas is much larger than the surface density of solids; when the metallicity $Z(r,t) = \Sigmap/\Sigma$ 
approaches or exceeds unity, solids begin to influence the evolution 
of the gas through angular momentum exchange and possible gravitational 
instabilities. Barring these cases, the standard evolution equation for 
$\Sigma(r,t)$ is
\begin{equation}
\frac{\partial \Sigma}{\partial t} +\frac{1}{r} \frac{\partial}{\partial r} 
\left( r u \Sigma \right) = - \dot{\Sigma}_{\rm w} \mbox{   ,   }
\label{eq:gasevol}
\end{equation}
where $u$ is the typical radial velocity of the gas required by conservation 
of angular momentum in the accretion disk,
\begin{equation}
u = - \frac{3}{ r^{1/2} \Sigma} \frac{\partial}{\partial r} \left( r^{1/2} 
\nut \Sigma \right) \mbox{   ,   }
\label{eq:ugas}
\end{equation}
and $\dot{\Sigma}_{\rm w}$ (where the dot from here on always 
denotes differentiation with respect to the time $t$) 
is the gas photo-evaporation rate modeled 
following the parametrization of AA07 (see Appendix A).

The gas turbulent diffusivity $\nut$ is modeled using the standard $\alpha$
-model
\begin{equation}
\nut = \alphat c h = \alphat \sqrt{\gamma} \Omega_{\rm K} h^2 \mbox{   ,   }
\end{equation}
where $c$ is the local sound speed and $\gamma$ is the adiabatic index 
of the gas. 
Note that there is a degeneracy between models with constant $\alphat$ and one 
particular temperature profile, and models with non-constant $\alphat$ and 
another temperature profile yielding the same value of $\nut$. This degeneracy 
combined with the crude $\alpha-$parametrization of turbulent transport used 
justifies the selection of a very simple temperature profile:
\begin{equation}
T_m(r) = \overline{T} \rau^q \mbox{   ,   }
\end{equation}
where $\rau$ is the distance to the central star in astronomical units. The 
scaleheight of the disk then varies as
\begin{equation}
h(r) = \overline{h} \rau^{(q+3)/2} \mbox{   .   }
\end{equation}

In what follows, I adopt the same disk model as that used by AA07: 
\begin{eqnarray}
&& q  = - 1/2 \mbox{   ,   }\nonumber \\
&& \hbarau = 0.0333 \mbox{   .   }
\label{eq:diskmodel}
\end{eqnarray}
Note that AA07 define $q$ as the power index of $h(r)$ instead of the power 
index of $T_m(r)$ used here; the apparently different values do correctly 
represent the same model. Although the numerical algorithm I have developed 
can be used with any input for $q$ and $\hbarau$, this particular value of $q$ 
is preferred as it greatly simplifies the analytical interpretation of the 
numerical results; indeed, in this case $\nut$ scales linearly with radius, 
a feature which turns out to be particularly useful.

\subsection{Evolution of vapor species}
\label{subsec:modelvap}

Chemical species in vapor form are evolved separately 
using the following standard advection-diffusion equation 
for a contaminant in a fluid of density $\Sigma$ moving with velocity $u$:
\begin{equation}
\frac{\partial \Sigma^i_{\rm v}}{\partial t} + \frac{1}{r} 
\frac{\partial}{\partial r}( r u \Sigma^i_{\rm v} ) = \frac{1}{r} 
\frac{\partial}{\partial r} \left[ r \nut \Sigma 
\frac{\partial}{\partial r} \left( \frac{\Sigma^i_{\rm v}}{\Sigma} \right) \right]\mbox{   ,   }
\label{eq:vaporevol}
\end{equation}
where it was implicitly assumed that the diffusivities of each chemical 
species are equal to the gas viscosity, and $u$ is given by equation 
(\ref{eq:ugas}). Sublimation and condensation are assumed to be instantaneous 
on the timescales considered and are calculated as a separate numerical step 
(see \S\ref{subsec:sublim}).
 
\subsection{Particle size distribution function}
\label{subsec:sizefunc} 

Collisional encounters between solid particles can result in their
coagulation or mutual shattering, the latter sometimes followed by the
re-accretion of material onto the largest remaining fragments. However
complex the mechanisms considered are, the size distribution function
of the particles is naturally expected to relax to a quasi-steady
equilibrium power-law within a few collision times. Theoretical
arguments on the steady-state nature of the collisional cascade imply
that the power-law index depends on the relationship between the
relative velocities of the objects and their material strengths 
(O'Brien \& Greenberg, 2003). Such power-laws are observed in the 
ISM (with index -3.5, Mathis, Rumpl \& Nordsieck, 1977), for
Kuiper-belt objects (with varying index depending on the size range)
and for asteroid-belt objects. This model is constructed by assuming
that encounters are frequent enough to maintain a quasi-steady
equilibrium, which results in a power-law size distribution (with
fixed index -3.5) for all particles of size less than $\smax$:
\begin{eqnarray}
&& \frac{\dd n}{\dd s} = \frac{\nmax}{\smax}\left(\frac{s}{\smax}\right)^{-3.5} 
\mbox{   for   } s \in [\smin,\smax] \mbox{   ,   } \nonumber \\
&& \frac{\dd n}{\dd s} = 0  \mbox{   otherwise  }
\label{eq:dnds}
\end{eqnarray}
where I allow the normalizing density $\nmax$, and the maximum particle 
size $\smax$ to vary both with radius and with time. The minimum
particle size $\smin$ is fixed, although its value does not influence 
the dynamical evolution of the disk as long as $\smax \gg \smin$ 
(since most of the solid mass is contained in the largest grains). Note
the value of $\smin$ influences the SED since the smallest grains 
contribute the most to the total emitting surface area. 

If the particles are spherical with uniform solid density $\rhos$ then 
the total density of solids is 
\begin{equation}
\rhop = \int_{\smin}^{\smax} \frac{\dd n}{\dd s} m(s) \dd s = 2 \nmax \mmax 
\end{equation}
provided $\smin \ll \smax$, where $m(s)$ is the mass of particles of
size $s$, and $\mmax$ is the mass of particles of size $\smax$ namely
\begin{equation}
\mmax = \frac{4\pi}{3} \rhos \smax^3 \mbox{   .   }
\end{equation}
This power-law size distribution function implies that 50\% of the total
mass is contained in particles of size $s \in [0.25
\smax,\smax]$.
 
The total surface density of particles is 
\begin{equation}
\Sigmap(r,t) = \sum_i  \Sigma^i_{\rm p}(r,t) \mbox{   .   }
\end{equation}
All condensed heavy elements present at a particular radius $r$ are assumed
to be fully mixed, or in other words, each particle has a mixed 
chemical composition that can vary depending on its radial position within 
the disk. Within this assumption, $\nmax$ can be related to the
{\it total} density of solids only, and within the particle disk (near the 
disk midplane), is directly related to the total surface density of particles 
via the equation
\begin{equation}
\nmax = \frac{\Sigmap}{2\mmax\sqrt{2\pi} \hp} 
\end{equation}
(assuming $\rhop$ has a Gaussian profile across the disk with 
scaleheight $\hp$). Note that the particle scaleheight $\hp$ 
depends on the mechanism 
exciting the intrinsic particle dispersion, which can be frictional coupling
with turbulent eddies or mutual gravitational interactions. It is naturally
independent of the particle species considered. 
Explicit expressions for $\hp$ in these two limits are given below.

\subsection{Particle motion}
\label{subsec:motion}

Motion of particles within the disk can be induced by various
possible forces: Brownian motion, motion induced by frictional 
drag with the gas and motion induced by interactions with the gravitational 
potential of the central star or that of other large planetesimals. 
The dominant term depends on the particle size.

Since the only particles 
considered here have size $\smax$, Brownian motion is typically
negligible. In a turbulent nebula, particles of
various sizes couple via gas drag to the turbulent eddies and can
acquire significant velocities when their typical stopping
time is comparable with the eddy turnover time. Larger particles are
only weakly coupled  with the gas but undergo significant
gravitational interactions with each other which constantly excite
their eccentricities and inclinations. These mechanisms can be thought
of as various kinds of stochastic forcing. Finally, non-stochastic forces
arise from the gravitational potential of the 
central star, and when combined with gas drag, can cause particles to sediment 
towards the mid-plane of the disk as well as spiral inward (occasionally 
outward).

These regimes are now described in more detail.

\subsubsection{Turbulence-induced dynamics}

In this section, I summarize existing results on the statistical properties 
of the dust dynamics resulting from their frictional coupling with turbulent 
eddies, and apply them to the problem at hand.\\
\\
{\it 1. Frictional drag.} Particles are coupled to the gas 
through frictional drag. The
amplitude of the drag force is statistically proportional to the
relative velocity between the particle and the gas, with a
proportionality constant that depends on whether the particle size is
smaller or larger than the mean-free-path of the gas molecules 
$\lambda_{\rm mfp}$ (Whipple, 1972).

If the particle is much smaller than $\lambda_{\rm mfp}$ (Epstein regime), 
drag forces originate from random collisions with the gas molecules, and the
typical timescale within which the particle will stop relative to the
gas is
\begin{equation}
\tau(s) = \frac{s\rhos}{\rho c}\mbox{   .   }
\label{eq:taueps} 
\end{equation} 

If the particle size is much larger than $\lambda_{\rm mfp}$ (Stokes regime) 
then the gas drag is principally caused by
the turbulent wake induced by the particles as it passes through the
gas. In this case, the particle stopping time is
\begin{equation}
\tau(s) = \frac{s\rhos}{\rho C_{\rm D} \sigma} \mbox{   ,   }
\label{eq:tausto} 
\end{equation}
where $\sigma$ is the typical velocity of the particle with respect to 
the gas, and the constant $C_{\rm D} \simeq 0.165$ (see Whipple 1972, 
Garaud, Barriere-Fouchet \& Lin 2004).

In what follows, it is useful to define $St(s)$ as the ratio 
of the local stopping time 
to the local orbital time $\taud = 2\pi /\Omega_{\rm K}$ (Weidenschilling, 1977), 
also called the Stokes number:
\begin{equation}
St(s) = \frac{\tau(s)}{\taud}  \mbox{   .   }
\end{equation}
Note that the Stokes number is equally as often defined as $\Omega_{\rm K}\tau(s)$ 
by other authors (Dullemond \& Dominik 2005 for instance). 
\\
\\
{\it 2. Relative velocities of particles.} As first estimated by 
Voelk {\it et al.} (1980) and summarized by
Dullemond \& Dominik (2005) (see also Weidenschilling
1984), particles of various sizes can acquire significant relative velocities 
through their frictional coupling with turbulent eddies.
This effect depends on the relative values of the eddy turnover time and
of the particle stopping time. For Kolmogorov turbulence with
large-scale eddy velocity $v_e \simeq \sqrt{\alphat} c$ and
large-scale turnover time comparable with the dynamical timescale
$\taud$, the Reynolds number $Re = \ve \taud^2/\nu$ determines the
eddy turnover time at the dissipation scale as $\tau_\nu = \taud
Re^{-1/2}$. Then, for two particles of respective stopping times
$\tau(s)$ and $\tau(s')$
\begin{eqnarray}
&& \Delta v(s,s') = \left[ \frac{(\tau(s)-\tau(s'))^2}{\taud
(\tau(s)+\tau(s'))}\right]^{1/2} \ve \mbox{   if   } \tau(s'),\tau(s)
\le \tau_\nu \mbox{   ,   } \\
&& \Delta v(s,s') = \ve \mbox{   if
} \tau(s') \le \taud \le \tau(s) \mbox{   ,   } \nonumber \\ 
&& \Delta v(s,s') =
\left[ \frac{\taud}{\taud + \tau(s)} + \frac{\taud}{\taud+\tau(s')}\right]^{1/2} \ve \mbox{   if   }
\taud \le \tau(s'), \tau(s) \mbox{   ,   } \nonumber \\ 
&& \Delta v(s,s') =
\frac{3}{\tau(s)+\tau(s')} \left[
\frac{\max(\tau(s),\tau(s'))^{3}}{\taud}\right]^{1/2} \ve \mbox{
otherwise.} \nonumber
\label{eq:relvel}
\end{eqnarray}
Note that in the first limit I have set $\sqrt{\ln Re /2} = 1$ for 
simplicity, which is underestimating the true collisional velocity 
by a factor of no more than about 4. This factor will be compensated 
for later (see \S\ref{subsec:growth}). Also note that the expression for the 
relative velocities in (\ref{eq:relvel}) has been corrected from 
that of Weidenschilling (1984) or Dullemond \& Dominik (2005) to account for 
an error pointed out by Ormel \& Cuzzi (2007).
\\
\\
{\it 3. Particle diffusion and effective Schmidt number.} 
The standard parametrization for the stochastic motion of particles of single 
size
$s$ coupled by gas drag to turbulent eddies is through the introduction of 
a turbulent diffusive mass 
flux $f_{\rm t}(s)$ in the particle continuity equation, typically
\begin{equation}
f_{\rm t}(s) = - \rho \Dp(s) \nabla \left(\frac{\rhop(s)}{\rho}  \right)\mbox{   ,   }
\label{eq:turbfluxs}
\end{equation}
where the {\it turbulent diffusivity} $\Dp(s)$ is related to
$\nut$ through the size-dependent Schmidt number
\begin{equation}
\Dp(s) = \frac{\nut}{Sc(s)}\mbox{   .   }
\end{equation}
The smallest particles are fully coupled with the gas so that $Sc(s)
\simeq 1$ if $\tau(s) \gg \taud$. The standard parametrization for
the Schmidt number in the case of large particles has long been $Sc(s)
\simeq St(s)$ (see for instance Dubrulle, Morfill \& Sterzik, 1995), 
so that $Sc(s)$ can be crudely approximated as $Sc(s) =
1 + St(s)$. Recent numerical and analytical work have shed
doubts on this  formula in favor of $Sc(s) \propto St^2(s)$ for large
particles (Carballido, Fromang \& Papaloizou 2006) and
have also questioned  the validity of equation
(\ref{eq:turbfluxs}) in favor of a  different formalism involving the
equilibrium solution a Fokker-Plank equation. Since these very
recent studies have not yet been fully completed (in particular, they
only consider particle diffusion in the $z$-direction and 
do not propose an alternative formalism  for the radial diffusion of
particles in the disk), I continue for the moment to adopt the 
standard parametrization of the Schmidt number $Sc(s) = 1+St(s)$.

For a fluid containing a size distribution of particles, 
the local diffusive mass flux of particles is obtained by 
integrating $f_{\rm t}(s)$ across all sizes, yielding
\begin{equation}
f_{\rm t} = - \Dp \rho \nabla\left(\frac{\rhop}{\rho}\right)\mbox{   ,   }
\label{eq:diffuflux}
\end{equation}
with $\Dp = \nut /Sc_{\rm eff}$ and the effective Schmidt number 
$Sc_{\rm eff}$ being
\begin{equation}
\Sceff = \frac{\sqrt{\Stmax}}{\arctan(\sqrt{\Stmax})}\mbox{   .   }
\label{eq:sceff}
\end{equation}
Note that $\Sceff$ is of order unity when $\Stmax \rightarrow 0$, 
as expected, while $\Sceff \simeq 2 \sqrt{\Stmax} / \pi$ if 
$\Stmax \rightarrow \infty$. This is quite different from the single 
particle size case, where the Schmidt number scales linearly with
particle size instead of with $\sqrt{\smax}$ in the decoupled limit. 
This reflects the fact that smaller particles remain well-coupled with 
the gas even when particles of size $\smax$ are fully decoupled.
\\
\\
{\it 4. Dust disk scaleheight.} Following the work of Dubrulle, 
Morfill \& Sterzik (1995), the dust 
disk scaleheight $\hp$ can be estimated by seeking stationary solutions 
of the settling/diffusion equation
\begin{equation}
\frac{\partial \rhop}{\partial t} - \frac{1}{3} \frac{\partial}{\partial z} 
\left( z\Omega_{\rm K}^2  \tau(\smax) \rhop \right) = \frac{\partial}{\partial z} 
\left[ \rho \Dp \frac{\partial}{\partial z} \left(\frac{\rhop}{\rho} \right)\right] \mbox{   ,   }
\end{equation}
where the factor of 1/3 arises from the mass-weighted integral of the 
settling velocities over the dust-size distribution function. Integrating 
this equation with height above the disk and assuming steady-state yields
\begin{equation}
\hp = h \left( 1 + \frac{2\pi}{3} \frac{\Stmax \Sceff}{\alphat \sqrt{\gamma}}\right)^{-1/2}\mbox{   ,   }
\end{equation}
where $h$ is the gas scaleheight.

\subsubsection{Gravitationally-induced motions}

As described by Kokubo \& Ida (2002), the typical velocity dispersion of a 
swarm of planetesimals (which is also equal to their typical relative 
velocities) can be deduced from the balance between gravitational 
excitation by the largest bodies, and damping by gas drag. The typical 
timescale for the excitation of the dispersion $\sigma(s)$ of planetesimals 
of size $s$ by protoplanets of size $\smax$ is given by equation (9) of 
Kokubo \& Ida (2002)
\begin{equation}
T_{\rm ex} = \frac{4 r^2 b <i^2(s)>^{1/2} \sigma(s)^3}{G^2 \mmax^2 \ln \Lambda}\mbox{   ,   }
\end{equation}
where $\ln \Lambda$ is the Coulomb logarithm, typically of the order of a 
few (here, I set $\ln \Lambda = 3$). The 
typical orbital separation $b$ of the emerging protoplanets is of 
the order of a few Hill radii (Kokubo \& Ida 2002):
\begin{equation}
b = \tilde{b} \rhill = 10 \left( \frac{2\mmax}{3 M_\star} \right)^{1/3} r\mbox{   ,   }
\end{equation}
where $\tilde{b} = 10$. The average inclination of the planetesimals 
$<i^2(s)>^{1/2}$ is assumed to be of the order of the average eccentricity, 
so that $<e^2(s)>^{1/2} = 2<i^2(s)>^{1/2}$. Finally, the random 
velocity of the planetesimals is also assumed to be related to their 
average eccentricity by 
\begin{equation}
\sigma(s) = <e^2(s)>^{1/2} v_{\rm K}\mbox{   .   }
\end{equation}
The timescale for damping of the typical inclination and eccentricity 
of the planetesimals is dictated by Stokes drag, namely
\begin{equation}
T_{\rm dp} = \frac{2 m(s)}{C_{\rm D} \pi s^2 \rho \sigma(s)}\mbox{   .   }
\end{equation}
Equating the two timescales yields the velocity dispersion for 
planetesimals of size $s$ in the presence of protoplanets of size $\smax$
\begin{equation}
\sigma(s) = \left(\frac{3}{2}\right)^{1/15} \left[ \frac{4\ln \Lambda}{3}
 \sqrt{\gamma}St(s)\frac{2\pi}{C_{\rm D} \tilde{b}} \frac{h}{r}\right]^{1/5} 
\left( \frac{\mmax}{M_\star}\right)^{1/3} v_{\rm K}\mbox{   .   }
\end{equation}
As Kokubo \& Ida (2002) found, this expression is only weakly dependent 
on the planetesimal size. If the gravitational perturbations are assumed 
to be statistically independent, then the relative velocities of the 
planetesimals are equal to their velocity dispersion. The weak dependence 
on size then implies that one can approximate the typical scaleheight of 
the planetesimals as
\begin{equation}
\hp \simeq <i^{2}(\smax)>^{1/2} r\mbox{   .   }
\end{equation}

\subsection{Particle growth}
\label{subsec:growth}

In the proposed model, the particle size distribution function is parametrized
with the power-law form given in equation (\ref{eq:dnds}), under the 
assumption that such power-law is naturally maintained as the quasi-steady 
state outcome of a coagulation/shattering balance. The normalization factor 
$\nmax$ is directly related to the total surface density of the dust $\Sigmap$,
while the maximum achievable size $\smax$ slowly grows in time as a result of 
occasionally successful coagulation events. 

Following this idea, I model the evolution equation for $\smax$ from the 
standard coagulation equation 
\begin{equation}
\frac{\dd \mmax}{\dd t} = \int_{\smin}^{\smax} \frac{\dd n}{\dd s}(s')
m(s') \Delta v(\smax,s') A(\smax,s') \epsilon \dd
s'
\label{eq:growth1}
\end{equation} 
where $\Delta v(\smax,s')$ is the average relative velocity between
particles of size $\smax$ and size $s'$, $A(\smax,s')$ is the
collisional cross-section of the two particles and $\epsilon$ is the
sticking probability of the two particles after the collision, or 
can be alternatively thought of as the average mass fraction of 
the impactor that sticks to the target after each collision. Note that in 
principle $\epsilon$ could depend on the collisional velocity, 
on the structure of the particles and on their size. In what
follows, the function $\epsilon$ will be chosen to be constant across
all sizes and relative velocities for simplicity. This approximation
is rather unsatisfactory, but merely mirrors insufficient knowledge about
the exact characteristics of the dust or larger particles. It can also be 
thought of as a weighted average of the true collisional efficiency across 
all size ranges and all possible impact velocities.

\subsubsection{Growth of particles in the turbulent regime}

For solid particles typically smaller than a few kilometers gravitational 
focusing is negligible (see below). Within this approximation, the 
collisional cross-section of two particles is reduced to the combined
 geometrical cross-section:
\begin{equation}
A(s,s') = \pi(s+s')^2\mbox{   .   }
\end{equation}

Using the expressions derived in \S2.4.1 for the relative velocities 
and the particle disk scaleheight, it is now possible to re-write equation 
(\ref{eq:growth1}) in a much simpler form. Three limits must first be 
considered: $\tau(\smax) \ll \tau_\nu$,  $\tau_\nu < \tau(\smax) < \taud$ 
and $\taud \ll \tau(\smax)$.  \\
\\
{\it Case 1:  $\tau(\smax) \ll \tau_\nu$}. 
In this case the particle growth is governed by 
\begin{equation}
\frac{\dd \smax}{\dd t} = \frac{\Sigmap}{8 \rhos} \sqrt{2\pi\gamma} \frac{h}{\hp} 
\sqrt{\alphat \Stmax} \frac{I_1}{\taud} \mbox{   ,   }
\end{equation}
where the integral $I_1$ is given by
\begin{equation} 
I_1 = \int_{\frac{\smin}{\smax}}^{1}  \epsilon x^{-0.5} (1+x)^{3/2} (1-x) \dd x  \mbox{   .   }
\end{equation}
Assuming that the sticking efficiency $\epsilon$ is constant, and that 
$\smin/\smax \ll 1$ the integral simplifies to $I_1 \simeq 1.8 \epsilon$. \\
\\
{\it Case 2:   $\tau_\nu < \tau(\smax) < \taud$}.
In this case,
\begin{equation}
\frac{\dd \smax}{\dd t} = \frac{\Sigmap}{8 \rhos} \sqrt{2\pi\gamma} 
\frac{h}{\hp} \sqrt{\alphat \Stmax} \frac{I_2}{\taud} \mbox{   ,   }
\end{equation}
where the integral $I_2$ is given by
\begin{equation} 
I_2 = \int_{\frac{\smin}{\smax}}^{1}  3 \epsilon x^{-0.5} (1+x) \dd x  \mbox{   .   }
\end{equation}
Under the same assumptions as in Case 1, $I_2 \simeq 8 \epsilon$. \\
\\
{\it Case 3: $\taud \ll \tau(\smax)$}.
This third case is slightly more complex, as the integral over particle 
sizes must be split between 
two bins, namely $\tau(s') < \taud$ and $\tau(s')> \taud$. This 
yields (in the limit considered)
\begin{equation}
\frac{\dd \smax}{\dd t} = \frac{\Sigmap}{8 \rhos} \sqrt{2\pi\gamma} 
\frac{h}{\hp} \sqrt{\frac{\alphat}{\Stmax}} \frac{I_3 + I_4}{\taud} 
\mbox{   ,   }
\end{equation}
where $I_3 \simeq 2 \epsilon$ and $I_4 \simeq 5\epsilon \Stmax^{-0.1}$.

For simplicity, the three cases can be combined into one formula only, namely
\begin{equation}
\frac{\dd \smax}{\dd t} = \frac{\Sigmap}{\rhos} \sqrt{2\pi\gamma} 
\frac{h}{\hp} \sqrt{\frac{\alphat\Stmax}{1+ 64\Stmax^2 (2+5 \Stmax^{-0.1})^{-2}}}
 \frac{\epsilon}{\taud} \mbox{   .   }
\label{eq:groturb}
\end{equation}
This expression overestimates the growth rate of the smallest particles 
(i.e. case 1) by a factor of about four. This error closely compensate 
for the factor of 4 underestimate in the collisional velocity of the 
smallest particles deliberately made in equation (\ref{eq:relvel}). The 
proposed expression recovers the formula for grain growth proposed by 
Stepinski \& Valageas (1997) within factors of order unity (see their 
equation (38)). 

\subsubsection{Growth of particles in the gravitationally dominated regime}

In this regime, the collisional cross-section is 
equal to the geometrical cross-section augmented by a gravitational 
focusing factor:
\begin{equation}
A(s,\smax) = \pi(s+\smax)^2(1+\Theta) \mbox{   where   } \Theta = 
\frac{2G \mmax}{\smax \sigma^2(s)}\mbox{   .   }
\end{equation}
When the Safronov number $\Theta$ is large, this expression simplifies to
\begin{equation}
A(s,\smax)\simeq \frac{2\pi G \mmax \smax}{\sigma^2(s)} 
\left(1 + \frac{s}{\smax}\right)^2\mbox{   .   }
\label{eq:gravA}
\end{equation} 

In addition, as particles grow larger in size, most of solid material 
becomes concentrated in fewer and fewer objects, until isolation mass 
is reached (all of the available material is contained in one object). 
In this work, I assume that the growing protoplanet can indeed accrete 
all the material available within the region of the disk centered on 
$r$ and of width equal to $\Delta r$ with 
\begin{equation}
\Delta r = \min ( \sqrt{A(\smax,\smax)}, \tilde{b} \rhill)  \mbox{   .   }
\end{equation}
In other words, the total surface density of material available for 
growth (excluding the mass contained in the growing protoplanet itself) is
\begin{equation}
\Sigmap - \frac{\mmax}{2\pi r \Delta r}\mbox{   .   }
\label{eq:Sreduced}
\end{equation}

Finally, using the expressions derived in \S2.4.2 for the 
particle velocity dispersion and for the disk scaleheight, the 
growth of the largest object is found to be governed by the equation 
\begin{equation}
\frac{\dd \smax}{\dd t} = \frac{\nmax \mmax}{3} \frac{2\pi G \smax^2}{\sigma(\smax)} I_5 \mbox{   ,   }
\end{equation}
where
\begin{equation}
I_5 = \int_{\smin/\smax}^1 \epsilon x^{-0.7} (1+x)^2 \dd x \simeq 5.3 \epsilon\mbox{   ,   }
\end{equation}
and $\nmax$ is reduced to include only the material available 
for growth (see equation (\ref{eq:Sreduced})) , 
\begin{equation}
2 \nmax \mmax = \frac{\Sigmap - \frac{\mmax}{2\pi r \Delta r}}{\sqrt{2\pi} \hp}
\end{equation}
so that 
\begin{equation}
\frac{\dd \smax}{\dd t} \simeq 1.77 \epsilon \frac{\Sigmap 
- \frac{\mmax}{2\pi r \Delta r}}{\sqrt{2\pi} \hp} \frac{\pi G \smax^2}{\sigma(\smax)} \mbox{  .   }
\label{eq:grograv}
\end{equation}

\subsubsection{Transition size}

The transition between the collisional regime dominated by turbulence 
and the collisional regime dominated by gravitational interactions is 
determined by the size for which the estimates of the velocity 
dispersion are equal, namely when 
\begin{equation}
<e^{2}(\smax)>^{1/2} v_{\rm K} = \frac{v_e}{\sqrt{\Stmax}}\mbox{   .   }
\label{eq:transize}
\end{equation}
Note that although this size depends on the surface density and temperature of 
the gas, and therefore on the position within the disk, it is typically 
of the order of a few kilometers. Beyond the transition size, 
the Safronov number is indeed found to be much larger than unity, 
justifying the use of the approximation $\Theta \gg 1$ in equation (\ref{eq:gravA}). 

\subsection{Evolution of the surface density of particles}
\label{subsec:sigmapevol}

The equation of evolution the surface density for each species condensed into 
solid particles is given by Takeuchi, Clarke \& Lin (2005) for instance, as
\begin{equation}
\frac{\partial \Sigmap^i}{\partial t} + \frac{1}{r} \frac{\partial}{\partial r} 
\left(rF^i_{\rm t} + r \Sigmap^i \up \right) = 0\mbox{   ,   }
\label{eq:dustevol}
\end{equation}
where $F^i_{\rm t}$ is the vertically integrated equivalent diffused mass 
flux cause by gas turbulence for each particle species (see equation (\ref{eq:diffuflux})) 
and $\up$ is the mass-weighted drift velocity of the particles resulting 
from gas drag. 

\begin{figure}[ht]
\epsscale{1}
\plotone{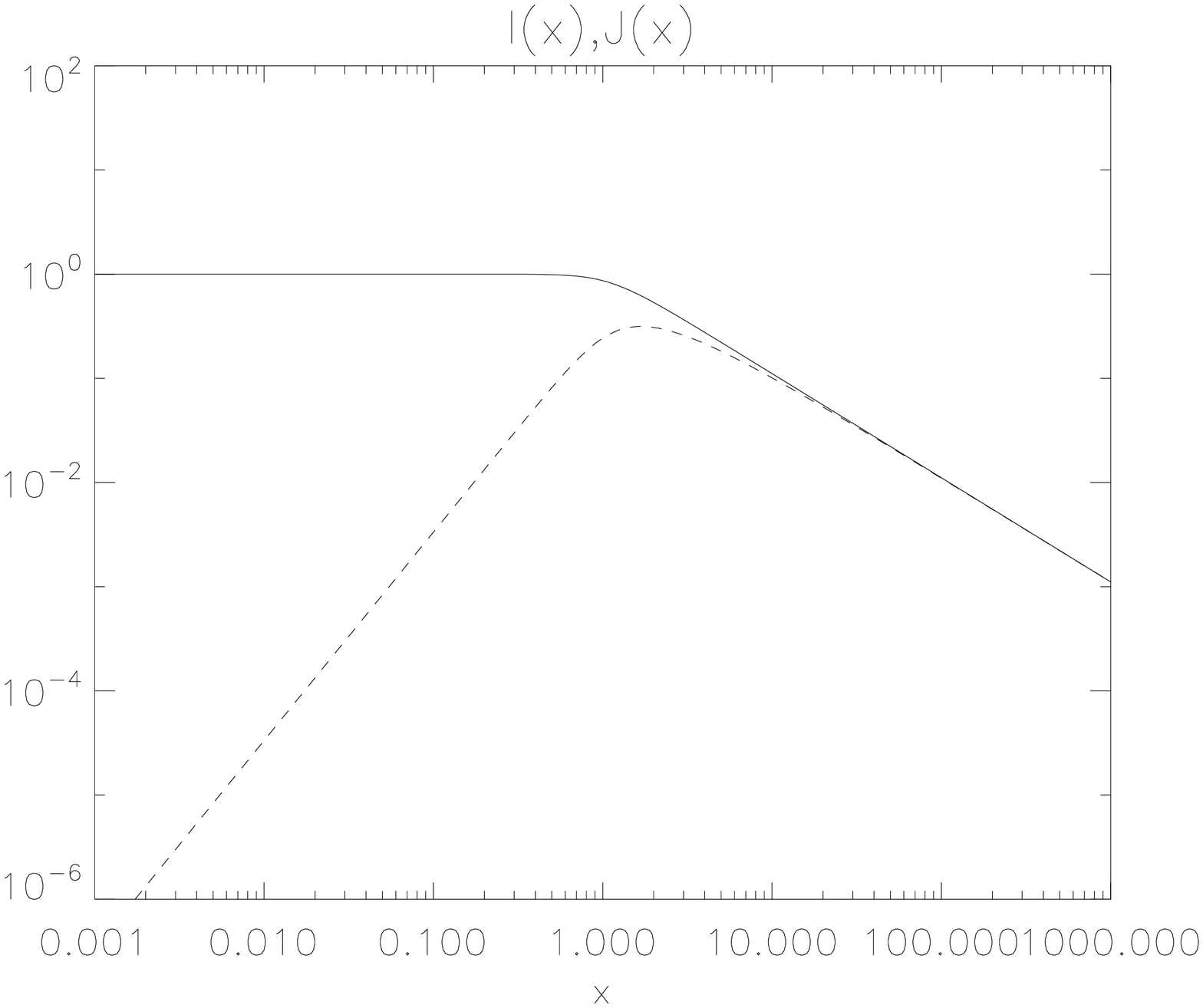}
\caption{I(x) (solid line) and J(x) (dashed line). As $x \rightarrow 0$, 
$I(x) \rightarrow 1$ and $J(x) \rightarrow   x^2/3$. As 
$x\rightarrow +\infty$, $I(x)$ and $J(x)$ both tend to $\sqrt{2}\pi/4x$.}
\label{fig:IJ}
\end{figure}

The radial velocity of a particle of size $s$ was calculated by Weidenschilling (1977) 
and can be written in the notation used here as 
\begin{equation}
\up(s) = \frac{u}{4\pi^2 St^2(s) + 1}   - 2\eta v_{\rm K}\frac{2\pi St(s)}{4\pi^2 St^2(s) + 1}\mbox{   ,   }
\label{eq:ups}
\end{equation}
where $\eta$ is related to the radial pressure gradient in the disk:
\begin{equation}
\eta = -\frac{1}{2} \frac{h^2}{r^2} \frac{\partial \ln p}{\partial \ln r} \mbox{   .   }
\label{eq:eta}
\end{equation}
Note that the constant $\eta$ reflects the difference between the 
typical orbital gas velocity and the Keplerian velocity at the same 
location in the disk. The mass-weighted average particle velocity is 
then determined by the integral
\begin{equation}
\up = \frac{\sqrt{2\pi} \hp}{\Sigmap} \int_{\smin}^{\smax} m(s) \up(s) \frac{\dd n}{\dd s} \dd s\mbox{   ,   }
\end{equation}
which integrates to
\begin{equation}
\up = u I(\sqrt{2\pi\Stmax}) - 2 \eta v_{\rm K} J(\sqrt{2\pi\Stmax})\mbox{   ,   }
\label{eq:upbar}
\end{equation}
where the functions $I$ and $J$ are given by 
\begin{eqnarray}
&& I(x) = \frac{\sqrt{2}}{4x} \left[ f_1(x) + f_2(x) \right] \mbox{   and   } \nonumber \\
&& J(x) = \frac{\sqrt{2}}{4x} \left[ - f_1(x) + f_2(x) \right] \mbox{   where   } \nonumber \\
&& f_1(x) = \frac{1}{2} \ln\left( \frac{x^2+x\sqrt{2}+1}{x^2-x\sqrt{2}+1}  \right) \mbox{   ,   } \nonumber \\
&& f_2(x) = \arctan(x\sqrt{2}+1) +\arctan(x\sqrt{2}-1) \mbox{   .   }
\end{eqnarray}
The functions $I$ and $J$ are shown in Figure \ref{fig:IJ}. Finally, note 
that planetary migration resulting from planet-disk 
interaction (type I or type II migration) is not taken into account here. 

\subsection{Sublimation/condensation}
\label{subsec:sublim}

Given the simplistic temperature profile used in this work, a simple 
sublimation/condensation model suffices.
The sublimation and condensation of each chemical species is assumed 
to be instantaneous in time. After
each timestep the new surface densities in solid and vapor 
forms are recalculated according to the very simple algorithm
\begin{eqnarray}
&& \Sigma^i(r,t) := \Sigmap^i(r,t) +\Sigma_{\rm v}^i(r,t) \mbox{   ,   } \nonumber \\
&& \Sigmap^i(r,t) := \frac{\Sigma^i(r,t)}{2}\left[ 1 + \tanh\left(\frac{T_i-T_m(r)}{\Delta T}\right) \right]\mbox{   ,   } \nonumber \\
&& \Sigma^i_{\rm v}(r,t) := \Sigma^i(r,t) - \Sigmap^i(r,t)\mbox{   ,   }
\end{eqnarray}
where $T_i$ is the typical sublimation temperature of the $i-$th 
species, and $\Delta T$ is taken to be 10K (in practise, the exact 
value of $\Delta T$ only influences the radial extent of the sublimation region).

\subsection{Numerical procedure}

The details of the numerical procedure adopted are given in Appendix
B, for reference. The algorithm constructed follows the simple pattern
at each timestep, from a given set of initial conditions; (i) test
whether particles of size $\smax$ are governed by turbulent or
gravitational interactions (ii) evolution of the particle size though
collisions using equations (\ref{eq:groturb}) or (\ref{eq:grograv}) 
accordingly (iii) evolution of the
gas density (iv) evolution of the vapor-phase of each species (v)
evolution of the particle phase of each species (vi)
condensation/sublimation and calculation the total surface density of
particles.

The numerical scheme adopted uses a standard split-operator
techniques, where diffusion terms are integrated using a
Crank-Nicholson algorithm, the advection terms are integrated using an
upwind explicit scheme and other nonlinear terms are integrated using
a 2nd order Adams-Bashforth scheme.

Depending on the spatial accuracy and the number of grain 
species studied, the typical integration time required to evolve of
a single disk over several Myr varies between a few hours and a day on a 
conventional desktop.

\section{Model parameters and initial conditions}
\label{sec:fiducial}

\subsection{Model parameters}

The numerical model requires a certain number of input parameters, 
listed in Table 1; these are separated between stellar parameters, 
photo-ionizing wind parameters, disk parameters and finally grain 
parameters. Default values for a ``fiducial model'' are also given. 
\\
\\
{\sc Table 1:} Fiducial model parameters \\
\vspace{0.05cm} \\
\begin{tabular}{|p{5cm}|p{0.5cm}|p{2cm}|}
\hline
Stellar Mass & $ M_\star $ & 1 $M_\odot$ \\
Stellar Luminosity & $L_\star$ & 1 $L_\odot$ \\
Stellar Radius & $ R_\star $ & 1 $R_\odot$ \\
Stellar Temperature & $T_\star$ & 1 $T_\odot$ \\
\hline
Sound speed of ionized gas & $ c_i$& $10^6$cm/s \\
Amplitude of photo-ionizing flux & $\Phi_i$ & $10^{42}$photons/s \\
\hline
Turbulent $\alpha$ & $\alphat $ & $10^{-2}$ \\
Scaleheight at 1AU & $\overline{h}_{\rm AU}$ & 0.0333\\
Temperature power law index & $q$ & -1/2 \\
Inner disk radius & $r_{\rm in}$ & 0.01 AU \\
Outer disk radius & $r_{\rm out}$ & 2000 AU \\
\hline
Solid density of grains & $\rho_s$ & $1.0$ \\
Sticking efficiency & $\epsilon$ & $10^{-2}$\\
Separation of protoplanets & $\tilde{b}$ & 10 \\
\hline
\end{tabular}
\\
\vspace{0.05cm} \\
\\
The various values selected for this fiducial model deserve comments. 
The star is 
chosen to be a solar-type star for ease of comparison of the results with 
the model of Stepinski \& Valageas (1997) and Ciesla \& Cuzzi (2006). Another 
possible choice would have been to select a typical T Tauri star 
($M_\star = 0.5 M_\odot$, $T_\star = 4000$K, and $R_\star = 2.5R_\odot$) 
which was done by Dullemond \& Dominik (2005). Detailed discussions on 
the values of the parameters associated with the photo-ionizing wind 
can be found in the work of AA07. 

The value of $\alphat$ is 
selected to be 0.01, which is a reasonable upper limit on the value that seems 
to be favored by  numerical simulations of MRI turbulence  
(Fromang \& Nelson 2006). However, by selecting a constant value of 
$\alphat$ both in time and space, I neglect possible effects of dead-zones 
(Gammie, 1996) which may not exist anyway (see Turner, Sano \& 
Dziourkevitch, 2007) 
as well as the transition from angular momentum 
transport dominated by gravitational instabilities to angular momentum 
transport dominated by MRI turbulence. The inner disk radius is chosen as a 
plausible location for the magnetospheric truncation radius 
(Hartmann, Hewett, \& Calvet, 1994)
while the outer disk radius is chosen at an arbitrarily large distance
 from the central star. 

The solid density of 
grains $\rhos$ is an elusive parameter since it is quite likely to 
vary strongly with time and with distance from the central star, both through 
repeated compaction events, self-gravity (in the case of large objects) 
and chemical composition. Here it is set to unity for simplicity, although 
this is admittedly not very satisfactory. The sticking efficiency is 
equally difficult to constrain a priori, although fascinating computational 
and experimental studies (see the review by Dominik {\it et al.}, 2007) 
are beginning to 
shed light on the subject. Here, I begin by assuming a value of 
0.01, and later discuss possible constraints on this value from 
observations of the grain surface density profile of disks.

\subsection{Model initial conditions}
\label{subsec:initcond}

The model described in this paper does not take into account the 
evolution of gas induced by self-gravity. It also ignores infall 
of mass onto the disk. As a consequence, it is limited to the 
study of disks which are gravitationally stable with negligible infall. 
The ``initial'' conditions should be thought of as the state of the 
disk after the Class I phase.

The required initial conditions of the model are: the initial surface 
density of the gas, the initial total surface density of heavy elements 
(both in gas and solid form), the respective proportion of heavy elements 
contained in each chemical species, and finally the initial maximum 
size $\smax$ of the dust particles. 

The initial surface density of the gas is selected to be a 
truncated power law (Clarke, Gendrin \& Sotomayor 2001)
\begin{equation}
\Sigma(r,0) = \frac{M_0}{2\pi R_0  r} e^{-r/R_0} \mbox{   ,   }
\end{equation}
and can therefore be easily characterized by the initial 
gas disk mass $M_0 = M(0)$ 
and the initial disk ``radius'' $R_0$. 
The initial total surface density of heavy elements 
(in both gas and solid form) is chosen to be a constant fraction of $\Sigma(r,0)$, with
\begin{equation}
\Sigmap(r,0) = Z_0  \Sigma(r,0) \mbox{   ,   }
\end{equation}
and thus can be characterized by one parameter only, 
namely the initial metallicity fraction $Z_0$.
The code is written in a very versatile way which allows the user to 
decide how many separate chemical elements to follow. The user needs to 
input the initial mass fraction of each chemical element, as well as their 
sublimation temperature under pressure and density conditions typical 
of accretion disks. As a first step, the sublimation/condensation routine is 
then run to decide what fraction of the total mass is in solid or in 
vapor form. The total solid particle density is then recalculated accordingly.

Finally, the initial size of the particles $\smax(r,0)$ must be chosen; 
for simplicity, it is assumed to be constant with $\smax(r,0) = s_{\rm max0}$. 
Although this is clearly an unrealistic initial condition, grain 
growth in the inner disk is so rapid that all ``memory'' of the 
initial conditions is lost within a few hundred years. On the other hand, since 
growth is negligible in the outer disk, $\smax(r,t) \simeq s_{\rm max0}$ there. 
Hence selecting the value of $s_{\rm max0}$ effectively determines the timescale 
for the evolution of solids in the disk (see \S\ref{subsec:retention}). 
While the fiducial model considers $s_{\rm max0}$ to be equal to the 
maximum plausible particle size in the MRN size-distribution function for the ISM, 
one could also imagine grains to grow even in the core-collapse phase. Suttner \&
Yorke (2001) found that grains could achieve sizes up to 10$\mu$m post-collapse, 
and so I will consider cases with varying initial conditions for $s_{\rm max0}$ 
in addition to the fiducial model (see \S\ref{subsec:mathsolids}).

Table 2 summarizes the initial condition input parameters, and gives typical 
values for a fiducial run.\\
\\
{\sc Table 2:} Fiducial model initial conditions.\\
\vspace{0.05cm} \\
\begin{tabular}{|p{3cm}|p{1.5cm}|p{2.5cm}|}
\hline
Initial disk mass & $M_0$ & $ 0.05 M_\star$ \\
Initial disk radius & $R_0$ & 30 AU \\
\hline
Initial metallicity & $ Z_0 $ &  $10^{-2}$ \\
Number of species & maxtype & 3 \\
Initial $\smax$ & $s_{\rm max0}$ & $1 \mu$m \\
\hline
\end{tabular}
\\
\vspace{0.05cm} \\
\\
The initial chemical composition of the dust, in the fiducial model,
is taken to be the following: 45\% ``Ices'' and other volatile
materials (with sublimation temperature $T_{\rm Ic} = 170$K), 35\%
refractory material (with sublimation temperature $T_{\rm Si} = 470$K)
and 20\% finally iron-based material (with sublimation temperature
$T_{\rm Fe} = 1300$K). The solid composition and sublimation 
temperatures are adapted
from Table 2 and Table 3 of Pollack {\it et al.} (1994) to account for a
reduced number of species.

\begin{figure}[ht]
\epsscale{1}
\plotone{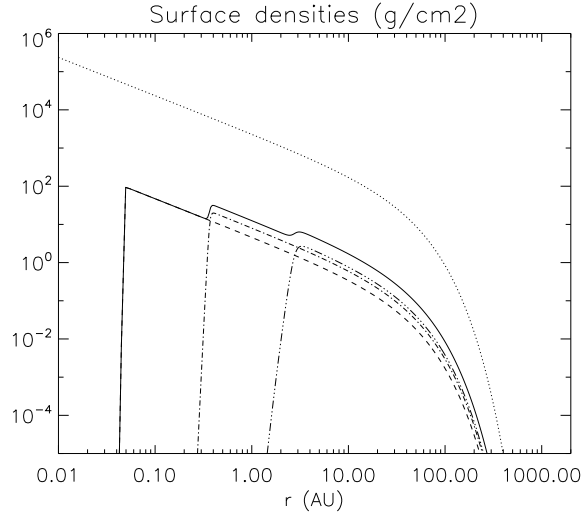}
\caption{Initial dust and gas surface densities in the fiducial disk model. 
The dotted line corresponds to the molecular gas and the solid line to the 
total surface density of solids. The three species considered are: the 
volatile material (dot-dot-dot-dash line), refractory material 
(dot-dash line) and the iron-rich material (dashed line).}
\label{fig:condinit}
\end{figure}

The fiducial initial model (after condensation/sublimation of the 
relevant species) is presented in Figure \ref{fig:condinit}.

\subsection{Model tests}

The numerical algorithm was tested against the results of 
AA07 for the evolution of the gas and grains
by using their initial conditions, switching off grain growth, 
sublimation and condensation, and by replacing equation (\ref{eq:upbar}) 
for the drift velocity with equation (\ref{eq:ups}). 
Both gas and grain evolution are found to be in perfect agreement, 
as required.

\section{Overview of results in the fiducial model}
\label{sec:results}

\begin{figure}[ht]
\epsscale{1.5}
\plotone{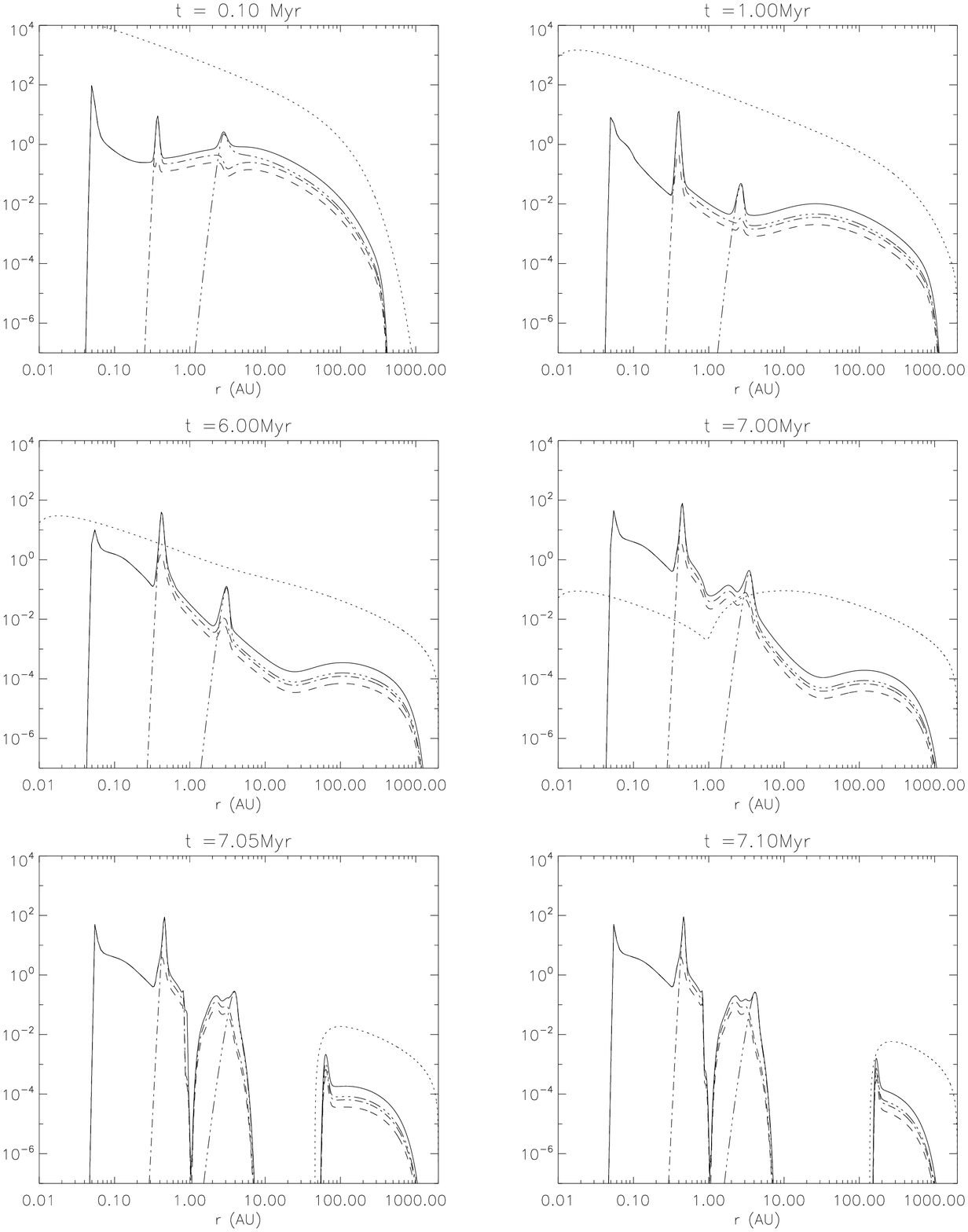}
\caption{Total surface density of gas (dotted line) and solids (solid line) at selected times. Also shown is the mass fraction in volatile materials (dot-dot-dot-dash line), refractory materials (dot-dash line) and iron-rich materials (dashed line).}
\label{fig:fiducialsd}
\end{figure}

\begin{figure}[ht]
\epsscale{2}
\plottwo{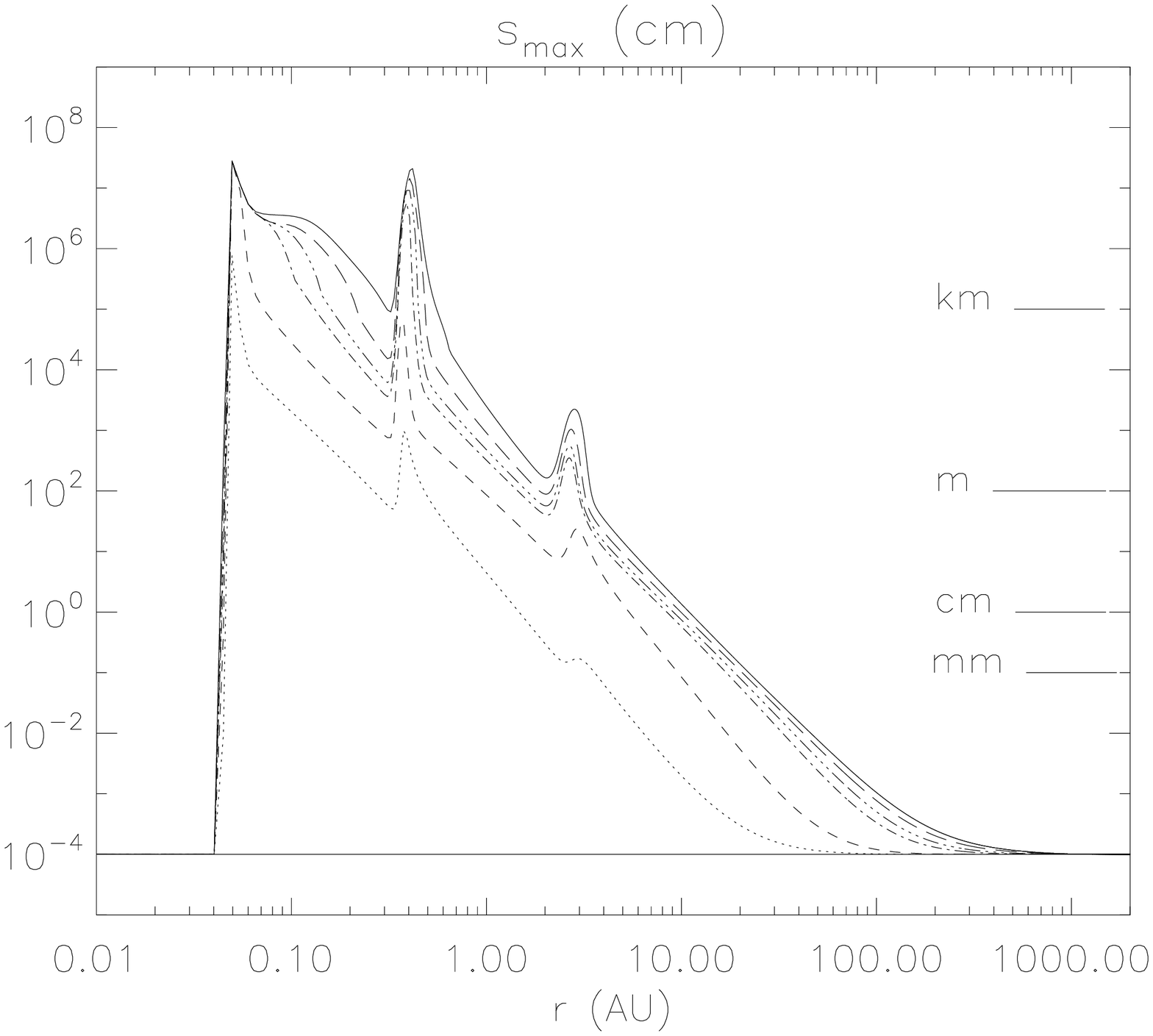}{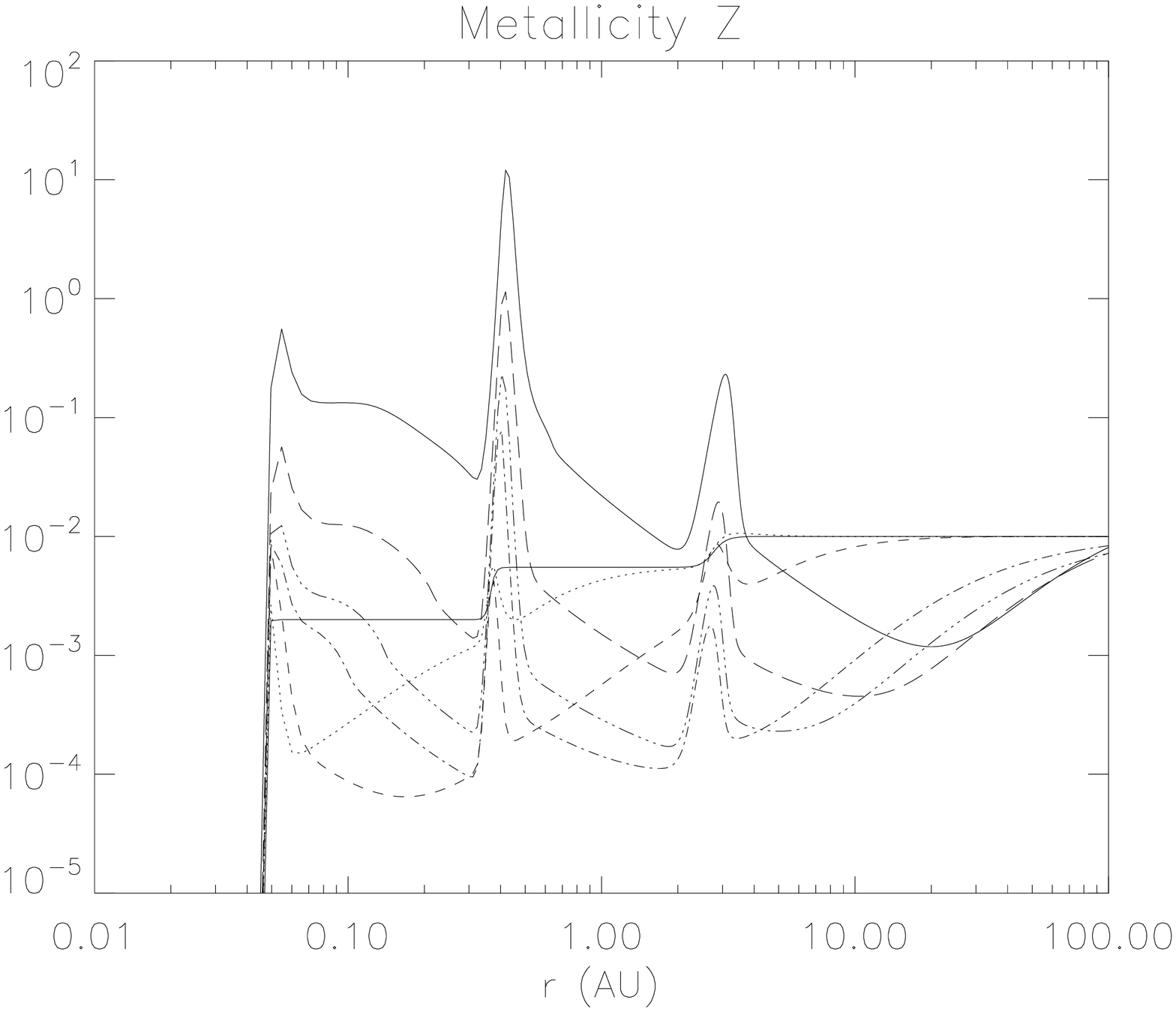}
\caption{Left: Evolution of the maximum particle size at selected times.
 From bottom to top, $t = $ 0, $10^4$ (dotted line), $10^5$ (dash line), 
$10^6$ (dot-dash line), $2\times 10^6$ (dot-dot-dot-dash line), $4\times 10^6$ 
(long-dash line) and $6\times 10^6$ yr (solid line). Note the strong growth 
peaks located near the respective sublimation lines, the plateau for 
$r < 0.1$AU where particles have reached isolation mass and the region 
of negligible growth for $r > 100$AU. Right: Metallicity fraction at 
the same selected times as in the left-hand-side figure. Note 
the strong initial reduction caused by the rapid inward drift of the 
particles, followed by gradual growth. The latter is caused by the 
reduction in $\Sigma$ rather than by an increase in $\Sigmap$.}
\label{fig:fiducialsmax}
\end{figure}

\begin{figure}[ht]
\epsscale{1}
\plotone{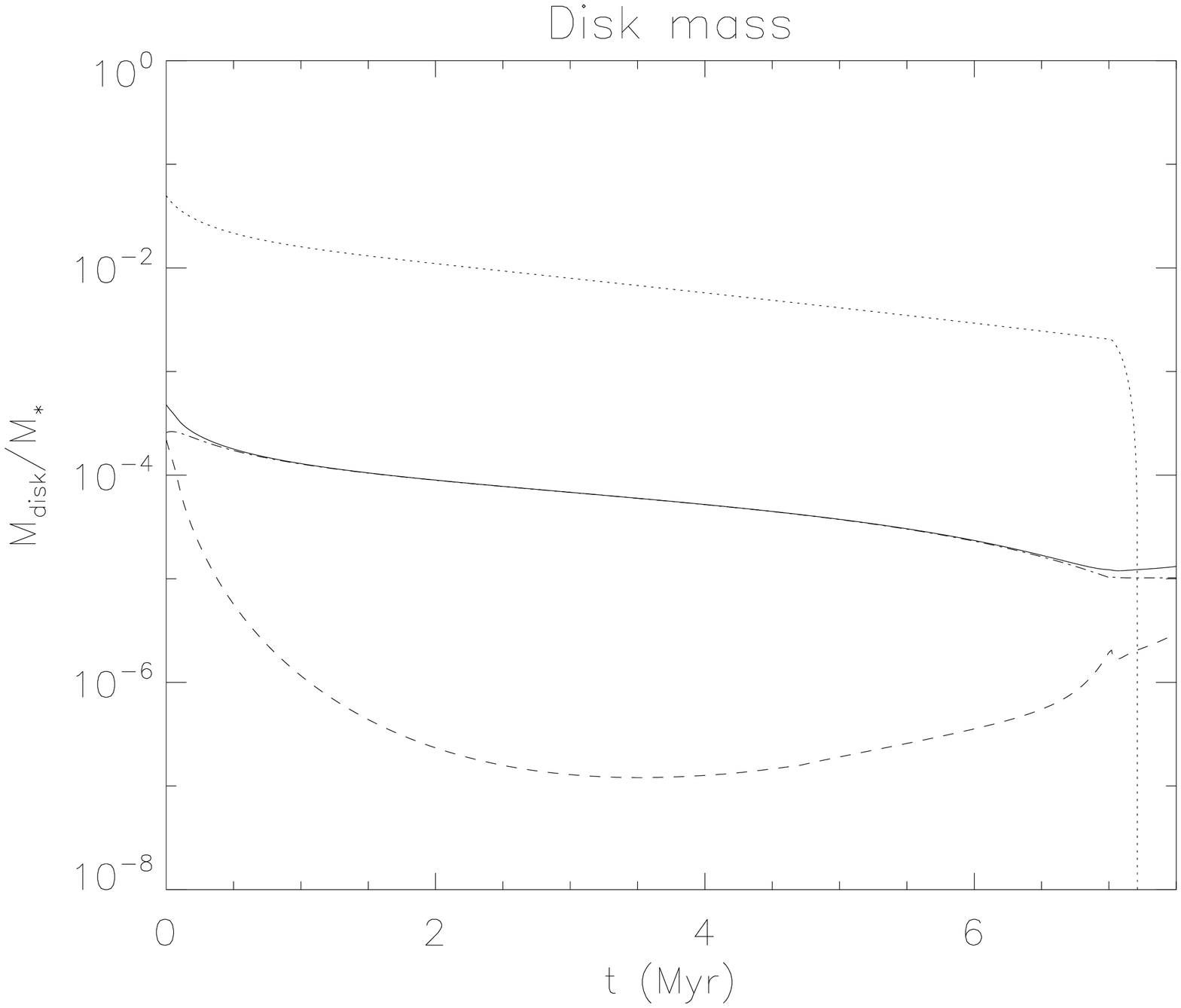}
\caption{Total disk mass in the fiducial model. The dotted line shows 
the integrated gas mass and the solid line shows the integrated solid 
mass. To illustrate the rapid loss of solids in the inner disk, the 
total disk mass contained in $r < 20$AU is shown in the dashed line, 
while the rest is shown in the dot-dash line.}
\label{fig:fiducialmass}
\end{figure}

The fiducial model presented in \S\ref{sec:fiducial} was integrated
forward in time until complete dispersal of the gas. Figure
\ref{fig:fiducialsd} shows the evolution of the surface density of the
gas, the total solid surface density as well as that of the three
species considered. Figures \ref{fig:fiducialsmax}a and 
\ref{fig:fiducialsmax}b show the evolution of the particle size and 
total metallicity as a function of radius and time. Finally, 
Figure \ref{fig:fiducialmass} shows the evolution in time of the total mass 
of gas and dust in the disk.

\subsection{Evolution of the gas surface density}
\label{subsec:fiducialsd}

The characteristic evolution of $\Sigma(r,t)$ under this 
particular photo-ionizing
wind  model has been extensively studied by Alexander, Clarke, \& Pringle, 
(2006a and 2006b) (see also Clarke, Gendrin \& Sotomayor, 2001). 
It can be seen in Figure \ref{fig:fiducialsd} as a dotted line, 
and in more detail in Figure \ref{fig:Sigmagas}. 
While the mass flux from photo-evaporation is negligible compared with the 
mass flux from viscous accretion/spreading, the disk undergoes a 
long period of near self-similar evolution. 
When both fluxes become comparable a depression appears in $\Sigma(r,t)$ and 
a gap eventually forms, here 
at radius $r_{\rm gap} = 0.9$ AU, at $t = 7$Myr. Within a few thousand years,
most of the gas in the inner disk has been
accreted onto the central star, while the radius of the hole begins
to expand as a result of direct photo-evaporation. 
At $t = 7.12$Myr, the hole radius has retreated to 200AU, and
finally beyond 500 AU after $t = 7.19$Myr. 

While the evolution of the gas is (in this model) independent of the 
evolution of solids, particle growth and particle migration are 
nonlinearly strongly coupled. 

\subsection{Particle growth}
\label{subsec:fiducialsmax}

The evolution of the maximum particle size $\smax(r,t)$ is shown in 
Figure \ref{fig:fiducialsmax}a both for very early times and at later times.
Grain growth is extremely rapid in the inner disk regions in the early
stages of disk evolution, in particular near sublimation lines.
Within just 100,000 yr, a characteristic shape to the curve $\smax(r,t)$
appears, which contains three different regions: (I) in the innermost
disk region ($r$ smaller than a fraction of 1 AU), a slightly tilted plateau 
corresponding to the particles having reached isolation mass; 
(II) a power-law region (from a fraction of 1 AU to about
100AU); (III) a region where grains have undergone negligible growth. 
Superimposed on this characteristic shape are a set of peaks
corresponding to the successive sublimation lines. The transition 
between region I and region II 
is easily identified as the transition between the gravitational regime 
and the turbulent regime; its steepness confirms that as soon as 
gravitational focusing 
becomes effective, the collision rate increases and particles rapidly 
reach isolation mass. The transition between region II and region III 
can also be easily identified as the region where the growth timescale of  
particles of size $s_{\rm max0}$ becomes comparable with the age of the disk.

Once established (after the first Myr), the global shape of the curve 
$\smax(r,t)$ varies little with
time (see solid lines), although particles within the sublimation 
region continue growing, and the three regions  
slowly expand outward.

\subsection{Solid density and chemical composition}
\label{subsec:fiducialsp}

The evolution of the solid density is shown in Figure \ref{fig:fiducialsd}.
Small particles well-coupled with the gas ($\Stmax \ll 1$)
closely follow its inward or outward motion depending on the 
radial position considered. As a result, during the initial viscous 
spreading of the disk (within the first Myr) 
a significant proportion of the mass in 
solids is transported outward with the gas creating a large reservoir 
of small dust grains at large radii ($r > 100$AU).
Meanwhile, particles in the inner regions of the disk ($r < 10$ AU)  
rapidly grow and begin to drift towards the central star 
differentially from the gas, which results in 
local changes in the metallicity $\Sigmap/\Sigma$. 
Figure \ref{fig:fiducialsmax}b shows the
evolution of metallicity in more detail, and reveals that the inner and 
intermediate disk go through  an initial phase of strong depletion 
in heavy elements. Later, the global evolution of the surface density of 
particles is controlled by the mass flux incoming from large radii. The 
observed increase in the metallicity is essentially related to the 
decrease in $\Sigma$ through photo-evaporation.

In addition to this global trend strong surface density peaks can be 
observed near the successive
sublimation lines. These are presumably caused by the differential 
drift of the solid and vapor form of each chemical species (Stepinski 
\& Valageas 1997, Ciesla \& Cuzzi, 2006). The peaks consistently stand roughly 
one to two orders of magnitude above the smoother ``background'' 
surface density profile, but do not appear to grow independently of it 
beyond the first 100,000 yrs. As the gas density decreases, 
the local metallicity near the sublimation lines 
steadily grows. By 4Myr, the sublimation line for refractory materials 
has equal content in gas and solids, suggesting the possibility of local 
onset of 
gravitational instability of solids (which is not modeled here). After 6Myr,
the two remaining sublimation lines (for the volatile and iron-rich material) 
also pass the same threshold. Interestingly, Figure \ref{fig:fiducialsd} 
reveals that the 
very rapid growth of material near each sublimation line traps a variety of
grain species into the growing bodies, so that the strong enhancement in 
the surface density of icy bodies near the volatile sublimation line is  
accompanied by an enhancement in the surface density of refractory materials 
and iron-rich materials. The 
same phenomenon is observed near the sublimation line for refractory materials.

\subsection{Disk masses}
\label{subsec:fiducialmass}

The evolution of the total mass in gas and solids in the disk is shown
in Figure \ref{fig:fiducialmass} as dotted and solid lines
respectively. Also shown are the total amount of solids found within
20 AU and outside of 20AU. At $t = 0$, the solid mass is equally
distributed between the inner ($<20$ AU) and the outer disk ($>20$AU);
Within a short time (of the order of 100,000yr), most of the mass 
in the inner disk accretes onto the central star, while the mass 
contained in the outer disk remains at a constant fraction of the 
disk mass in gas. Beyond this point, the mass in the inner disk is 
controlled by the flux of material drifting in from the outer disk.

When the gap opens (at 7Myr), the total mass of gas drops precipitously 
(within about 200,000 yr) while the total mass of solids remains constant. 
The total content of
solids left in the disk after complete photo-evaporation of the gas is
about 1.3 $\times 10^{-5} M_{\odot}$, or in other words about 4 Earth
masses. Only 20\% of this amount is located within the inner 20AU of
the disk, while the remaining 80\% are swept out to the outer disk.

\section{Mathematical interpretation of the results}
\label{sec:math}

In order to gain more insight into the numerical results for the 
fiducial model, it is useful to characterize and when possible quantify
some of the generic behaviors observed in the solutions. 

In this section I present both existing and new analytical results on the 
evolution of gas and solids in viscously evolving disks. While the complexity
of the system clearly precludes the existence of a closed-form 
fully analytical solution, there are certain limits where analytical 
efforts pay off. By comparing the analytical estimates derived with 
the exact outcomes of the numerical algorithm, I am able to test 
the numerical results on a systematic basis, and at the same time obtain 
strict constraints on the regimes of validity of the analytical solutions. 

Given that the evolution of the gas is more-or-less independent of the 
evolution of solids, much progress has already been done in describing
it analytically. These are presented in \S\ref{subsec:mathgas}. New results on the 
evolution of solids are presented in \S\ref{subsec:mathgrow} and \S\ref{subsec:mathsolids}.
\subsection{Evolution of the gas}
\label{subsec:mathgas}

\begin{figure}[ht]
\epsscale{1}
\plotone{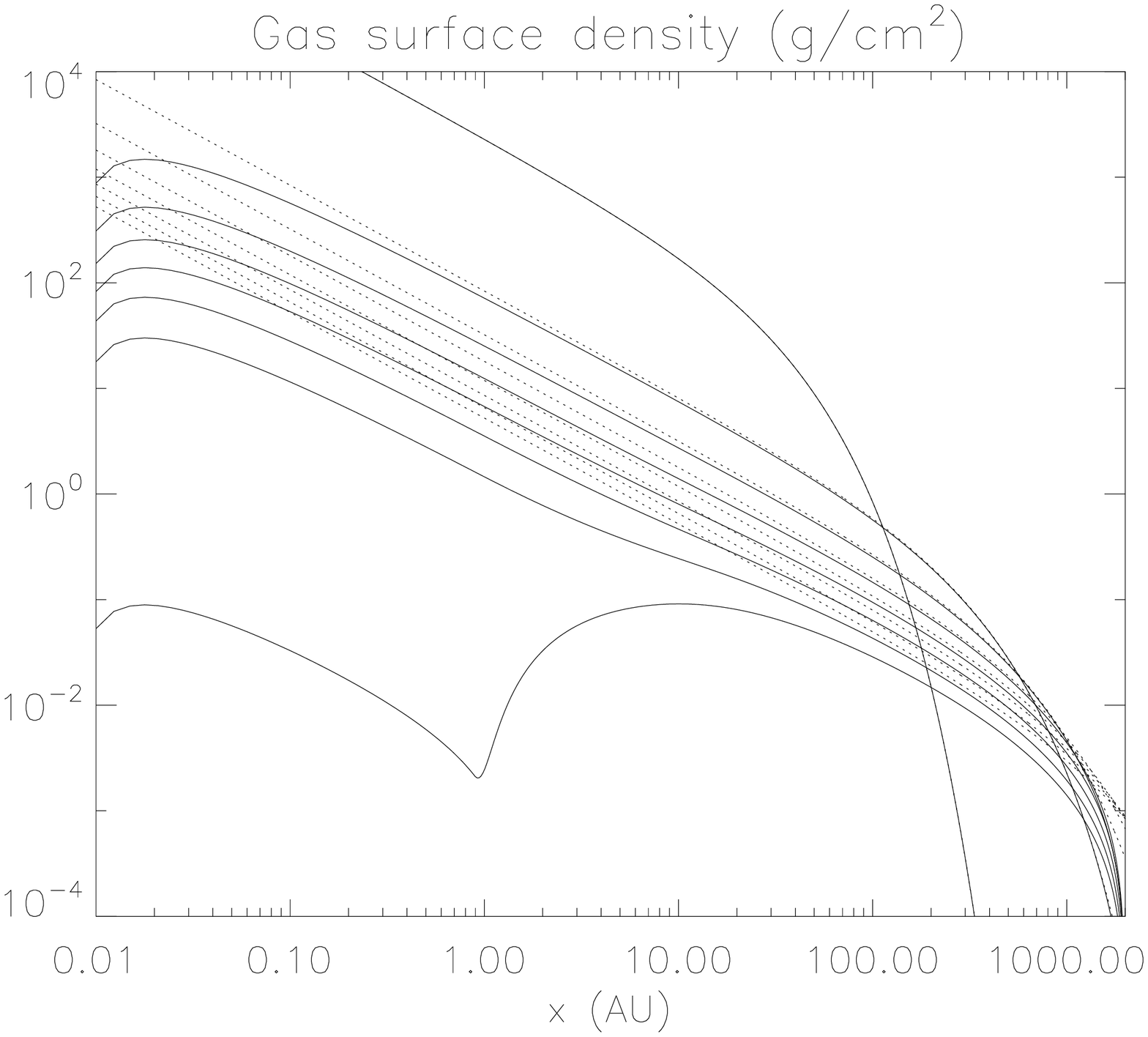}
\caption{Gas surface density in the fiducial model 
(from the top to the bottom curve) at $t$ = 0, 1, 2, 3, 4, 5, 6 and 7 Myr. 
The solid lines show the exact numerical solution, while the dotted lines
 show the analytical estimate provided by the self-similar solution 
(\ref{eq:approxsigma}) at the same times.}
\label{fig:Sigmagas}
\end{figure}

The evolution of the gas density is shown in more detail in Figure 
\ref{fig:Sigmagas}. Lynden-Bell \& Pringle (1974) 
(see also Hartmann {\it et al.} 1998)
showed that provided (i) the mass accretion 
rate due to viscous transport is much larger than the 
wind photo-evaporation rate and (ii) the disk is allowed to spread 
to infinity, then there exist a simple self-similar solution for $\Sigma(r,t)$:
\begin{eqnarray}
\Sigma(r,t) &=& \frac{M_0}{2\pi r R_0} T^{-3/2} \exp\left(-\frac{r}{R_0T}\right) 
\mbox{   with    } \nonumber \\
T &=& \frac{t}{\tau_{\rm v}} + 1\mbox{   ,   }
\label{eq:approxsigma}
\end{eqnarray}
where the viscous spreading time of the disk is 
$\tau_{\rm v} = R_0^2/3\nut(R_0)$. In this solution, the gas velocity 
is equal to 
\begin{equation}
u = - 3\nut(r) \left(\frac{1}{2r} - \frac{1}{R_0 T}\right)
\label{eq:ugasapprox}
\end{equation}
showing that $u < 0$ for $r < r_{\rm v}(t)$ (viscous accretion) 
and $u> 0$ for $r > r_{\rm v}(t)$ (viscous spreading), the critical 
radius being $r_{\rm v}(t) = R_0 T/2$. 
Figure \ref{fig:Sigmagas} compares the self-similar solution compares 
with the true numerical solution, and reveals excellent agreement at
early times, gradually deteriorating in the inner disk as time evolves 
and photo-evaporation 
begins to dominate the gas dynamics. In the outer disk 
the agreement remains globally much better 
(since the photo-evaporation rate is very low at large radii) and deteriorates
only slightly (by a factor of no more than a few) as expected when the 
critical radius reaches the outer boundary. Note that a much better
 approximation to $\Sigma(r,t)$ including the effects of photo-evaporation 
has been obtained by Ruden (2004).

In the early self-similar phase, and within $r_{\rm v}$, 
the mass accretion rate $\dot{M}(r,t) = 2\pi r u \Sigma$ 
is roughly constant with radius. The total gas disk mass decays 
as $M(t) = M_0 T^{-1/2}$ (neglecting the effects of the outer 
boundary condition), so that the gas accretion timescale increases 
linearly with the reduced time: $M/|\dot{M}| \simeq 2 T \tau_{\rm v}$. 

When the photo-evaporation rate becomes 
comparable with the accretion rate, a gap opens in the disk. 
Hollenbach {\it et al.} (1994) argued that the gap opening 
radius $r_{\rm gap}$ is located close to the gravitational 
radius $r_{\rm g} = G M_\star/c_i$, while Liffman (2003) and 
Font {\it et al.} (2004) 
revised this estimate to be a fraction of $r_{\rm g}$. Since 
$r_{\rm g}$ scales linearly 
with stellar mass so does $r_{\rm gap}$; in what follows, I adopt
\begin{equation}
r_{\rm gap} = 1 \left(\frac{M_\star}{M_\odot}\right) {\rm AU} \mbox{   .   }
\end{equation}
The time at which the gap opens
can be estimated by equating the wind mass loss rate $\dot{M}_{\rm w}$ 
to the viscous accretion rate in the self-similar solution 
(Clarke, Gendrin \& Sotomayor, 2001); this yields (provided $\tau_{\rm gap} \gg \tau_{\rm v}$)
\begin{equation}
\tau_{\rm gap} = \tau_{\rm v} \left( \frac{M_0}{2\tau_{\rm v} \dot{M}_{\rm w}} \right)^{2/3}\mbox{   .   }
\label{eq:taugap}
\end{equation}
The wind mass loss-rate prior to gap opening for the fiducial model was 
calculated by AA07 to be 
\begin{equation}
\dot{M}_{\rm w} \simeq 4 \times 10^{-10} M_\odot \mbox{/yr}\mbox{   .   }
\end{equation}
Table 3 compares the estimate from equation (\ref{eq:taugap}) 
with the outcome of numerical simulations with varying $R_0$ and $M_0$, 
showing that it is indeed a very good estimate for the gap formation 
timescale except for the fiducial model. This is because the actual viscous 
mass accretion rate 
in the numerical solution of the fiducial model deviates from the simple 
estimate of $M_0 T^{-3/2}/2\tau_{\rm v} $ at large times, when the disk spreads all 
the way  out to the outer edge of the numerical mesh\footnote{To get a 
better estimate for the gap formation time in this case, one should simply
not neglect the outer boundary term in the calculation of the total mass of 
the disk, see Hartmann {\it et al.} 1998.}.\\
\\
{\sc Table 3:} Gap opening time, actual and predicted\\
\vspace{0.05cm} \\
\begin{tabular}{|p{1.5cm}|p{1,8cm}|p{1.8cm}|p{1.8cm}|}
\hline
$R_0$ (AU) & $M_0 / M_\odot$ & $ \tau^{\rm num}_{\rm gap} $ (Myr) & $ \tau^{\rm pred}_{\rm gap} $ (Myr) \\
\hline
30 & 0.05 & 7.01 & 7.79 \\
10 & 0.05 & 5.35 & 5.40 \\
5 & 0.05 & 4.36 & 4.29 \\
10 & 0.01 & 1.96 & 1.84 \\
\hline
\end{tabular}
\\
\vspace{0.05cm} \\
\\
After the gap formation, supply of material from the outer disk is shut 
off and the inner disk rapidly clears of all gas. The gas clearing 
time can be estimated from the viscous timescale at the gap formation radius: 
\begin{equation}
\tau_{\rm clear} = \tau_{\rm v}\left(\frac{r_{\rm gap}}{R_0}\right)\mbox{   .   }
\end{equation} 
In the fiducial model, the gas clearing time is of the order of 4,000 yr 
only, and can be considered to be near-instantaneous. This is indeed seen in the
simulations (see Figure \ref{fig:fiducialsd}).

Direct photo-ionization of the gas at the hole edge results in a sharp 
change in the gas mass loss rate (see Appendix A). The evolution of the 
size of the hole can be derived from the work of Alexander, Clarke \& Pringle 
(2006b) 
to be
\begin{equation}
r_{\rm hole}(t) = r_{\rm gap} \left( 
\frac{t-\tau_{\rm gap}}{2\tau_{\rm clear}} + 1 \right)^2\mbox{   ,   }
\end{equation} 
which is again a good estimate of the hole radius derived from the numerical 
solution (see \S\ref{subsec:fiducialsd}) for $t > \tau_{\rm gap}$. 

\subsection{Grain growth}
\label{subsec:mathgrow}

\begin{figure}[ht]
\epsscale{1}
\plotone{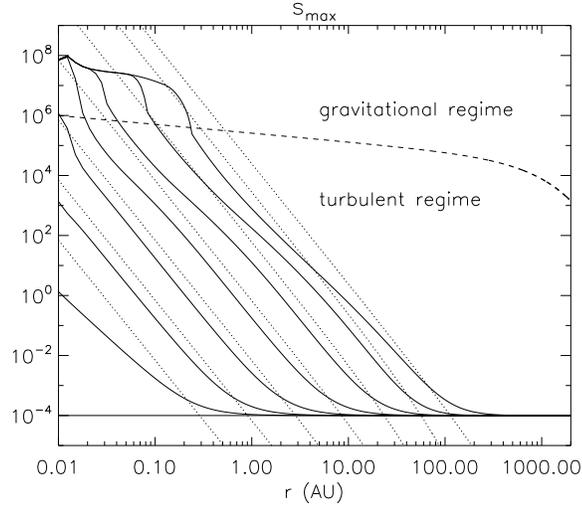}
\caption{Maximum grain size $\smax(r,t)$ at $t = 0$, 1, 10, 100, 1000, 
$10^4$, $10^5$ and finally $10^6$ yrs (solid lines, from bottom to top) 
obtained from the numerical simulation of the fiducial model with no 
sublimation/condensation. The dotted lines show the analytical estimates 
of equation (\ref{eq:approxsmax}) at the same times. The dashed line 
marks the transition between the turbulent and gravitational regime at $t = 1$Myr.}
\label{fig:smaxnosublim}
\end{figure}

For simplicity, in this section I focus on deriving 
analytical estimates for grain growth in disks  
where sublimation and condensation are neglected. 
For ease of comparison with the numerical model, an additional 
run was performed using the fiducial model but with no sublimation 
or condensation of material, yielding the 
solution shown on Figure \ref{fig:smaxnosublim}. After the 
initial rapid growth period (for t > $10^4$ yrs), one can note very clearly
the three regions of interest described earlier: region III where 
$\smax(r,t) \simeq s_{\rm max0}$; region II where 
$\smax(r,t)$ appears to follow  roughly a 
power law in $r$, and region I where the particles have reached 
isolation mass and stopped growing. In this figure the peaks 
associated with sublimation lines are naturally absent.

Despite the apparently simple behavior of the numerical solutions, modeling 
this evolution is a complex problem: the particle growth rate 
depends on $\Sigmap(r,t)$, which is regulated by 
the drift velocity of the particles, which depends on 
the particle size $\smax(r,t)$. The implied nonlinearities
preclude the existence of analytical solutions in most cases.

However, there exist a limit in which insight can be gained from simple
models: when the particle drift time is much longer than the particle 
growth time, one can expect the metallicity to remain close to its initial 
value, namely $\Sigmap/\Sigma = Z_0$. This happens in the very early stages 
of disk evolution (the limit of small $t$), as well as in the outer regions 
of the disk (region III) at all times. 

\subsubsection{Growth timescales}

To interpret the numerical solutions I consider the growth regime 
dominated by turbulent encounters, where particles of 
size $\smax$ follow the growth law given by equation (\ref{eq:groturb}).
The growth timescale of the particles is given by 
\begin{equation}
\tau_{\rm g} = \left(\frac{1}{\smax} \frac{\dd \smax}{\dd t} \right)^{-1} \mbox{   .   }
\label{eq:taugrow}
\end{equation}
In the limit where $\Stmax \ll 1$, the growth timescale is equal to 
\begin{equation}
\tau_{\rm g} = \frac{\Sigma}{\Sigma_{\rm p}} \sqrt{\frac{\Stmax}{\alphat}} 
\frac{\tau_{\rm d}}{\epsilon}\mbox{   ,   }
\label{eq:taugrowsmall}
\end{equation}
while in the limit where $\Stmax \gg 1$
\begin{equation}
\tau_{\rm g} =  2\sqrt{3} \gamma^{1/4} \frac{\Sigma}{\Sigma_{\rm p}} 
\Stmax^{3/4} \frac{\tau_{\rm d}}{\epsilon}\mbox{   .   }
\label{eq:taugrowlarge}
\end{equation}
Note that the second expression is independent of $\alphat$. This is 
related to the fact that for larger particles, coupling with the 
turbulent eddies is weakened, which reduces their relative velocities 
but also increases particle concentration through sedimentation. 

\begin{figure}[ht]
\epsscale{1}
\plotone{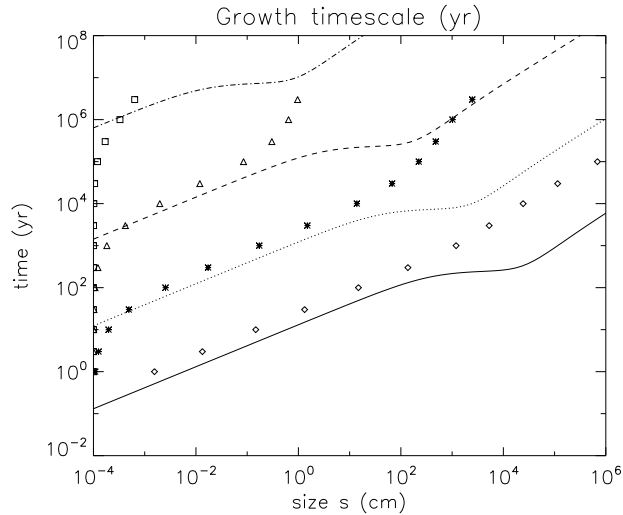}
\caption{Growth timescale for particles of size $s$ for the initial 
conditions of the fiducial disk described in \S\ref{subsec:initcond} with 
no sublimation/condensation; the curves show the growth timescale at 
$r $ = 0.1AU (solid line), $r =$ 1AU (dotted line), $r =$ 10AU 
(dashed line) and finally $r =$ 100AU (dot-dash line). The symbols 
show the true $\smax(r,t)$ at the same radii (0.1 AU: diamonds; 1 AU: 
stars; 10 AU: triangles; 100 AU: squares).}
\label{fig:growthtime}
\end{figure}

Figure 
\ref{fig:growthtime} shows (as lines) the exact analytical expression 
for the growth timescale obtained by combining equation (\ref{eq:taugrow}) 
with equation (\ref{eq:groturb}), and considering $\Sigmap/\Sigma$ 
fixed and equal to the initial metallicity $Z_0 = 0.01$. 
Growth timescales of particles at radii 0.1, 1, 10 and 100 AU are shown. 
The power laws seen in 
either limits are well-approximated by equations (\ref{eq:taugrowlarge}) 
and (\ref{eq:taugrowsmall}), and the flattening of the four curves 
corresponds to the transition $\Stmax \ll 1$ to $\Stmax \gg 1$. 
This figure is particularly useful for reading directly 
the maximum size of particles achievable at a given age in the disk should the
surface densities of dust and gas indeed remain constant over that period. For 
example by looking at the intersection of a horizontal line at $10^4$yr 
with each curve, one deduces that in a $10^4$yr-old disk there will be no 
significant growth beyond 
20 AU, 10-20$\mu$m-size particles at 10AU, m-size objects at 1 AU and finally, 
gravitationally dominated growth (towards isolation mass) below 0.1AU. Since  
$\tau_{\rm g} \propto 1/ Z_0\epsilon$, the growth timescale and maximum 
particle sizes for disks with other
values of the metallicity $Z_0$ or sticking efficiency $\epsilon$ 
can be read by translating the curves up or down accordingly.

In reality, the metallicity $Z = \Sigmap/\Sigma$ is of course {\it not} 
constant. Figure \ref{fig:growthtime} also shows 
(as symbols) the actual particle size achieved in the fiducial disk with no 
sublimation/condensation at the same selected radii. 
The lines and the symbols follow each other reasonably well in the 
expected limits (large radii or short times) up to
the point where $\Stmax \simeq 1$ at which point the simple analytical 
estimate systematically breaks down\footnote{The reason why
 $\Stmax \simeq 1$ is equivalent to the point where $Z$ begins to 
differ significantly from $Z_0$ is related to the fact most of the mass 
is contained in the particles of size $\smax$, which also happen to be
the particles with the highest inward drift velocity.}.

\subsubsection{Particle size}

Given the good fit found for particles with $\Stmax \ll 1$, 
I now approximate $\smax(r,t)$ by the value of the 
grain size for which the growth timescale equals the age of the disk: 
setting $\tau_{\rm g} = t$ in equation yields
(\ref{eq:taugrowsmall}) 
\begin{equation}
\smax(r,t) = \frac{\sqrt{2\pi\gamma}}{\rhos} \left(\frac{t}{{\rm 1 year}}\right)^2 
\left(\frac{\Sigmap}{\Sigma}\right)^2 \epsilon^2 \alpha_t \Sigma(r,t) r_{\rm AU}^{-3}  
\left(\frac{M_\star}{M_\odot}\right) \mbox{   .   }
\label{eq:approxsmax}
\end{equation}
If one assumes as before that $\Sigmap/\Sigma \simeq Z_0$ then 
one can get a rough estimate of $\smax(r,t)$ by combining 
equations (\ref{eq:approxsmax}) and (\ref{eq:approxsigma}), 
and shown in Figure \ref{fig:smaxnosublim} as dotted lines. 

As expected, the estimate for $\smax(r,t)$ is in fairly good a
agreement with the numerical results for small times; it correctly 
predicts the power law structure of the whole intermediate region 
(for $s < 1$km, roughly) at early times (t < $10^4$yr), but not 
so at later times (where the analytically predicted power law is 
too steep compared with the numerical results). This is again 
related to the fact that the surface density of particles becomes
 significantly depleted at later times.

The estimate for $\smax(r,t)$ also correctly predicts the transition 
between regions II and III of the disk. Since particle growth is 
fundamental to our understanding of disk SEDs, I now 
give an analytical estimate for this transition radius:
for early times (for $t < \tau_{\rm v}$), $r^{II}_{III}(t)$ is given by 
\begin{equation}
r^{\rm II}_{\rm III}(t) =  \left[ \frac{Z_0^2 \epsilon^2 \alphat}{St_{\rm max0}} 
\frac{R_0}{1 {\rm AU}}\right]^{1/4} \left(\frac{t}{1{\rm year}}\right)^{1/2} 
\left(\frac{M_\star}{M_\odot}\right)^{1/4}{\rm AU}\mbox{   ,   }
\end{equation}
where 
\begin{equation}
St_{\rm max0} = \frac{2\pi R_0^2 s_{\rm max0} \rhos }{\sqrt{2\pi \gamma} M_0}
\label{eq:stmax0}
\end{equation}
is the Stokes number at $t=0$ and $r = R_0$ of particles of size $s_{\rm max0}$. 
For later times ($ t > \tau_{\rm v}$)
\begin{equation}
r^{\rm II}_{\rm III}(t) = \left[ \frac{Z_0^2 \epsilon^2 \alphat}{St_{\rm max0}} 
\frac{R_0}{1 {\rm AU}} \right]^{1/4}  \left(\frac{\tau_{\rm v}}{1 {\rm year}}\right)^{3/8}   
\left(\frac{t}{1 {\rm year}}\right)^{1/8}  \left(\frac{M_\star}{M_\odot}\right)^{1/4} {\rm AU}\mbox{   .   }
\end{equation}

\subsubsection{Gravitational regime} 

The transition from the turbulent regime to the gravitational regime 
(region II to region I) is easily understood by considering equation 
(\ref{eq:transize}). The transition size $\tilde{s}_{\rm max}$ is given by 
\begin{equation}
\tilde{s}_{\rm smax}(r,t) = 7.1 \left(\frac{M_\star}{M_\sun}\right)^{10/51} 
\left(\frac{h}{r}\right)^{8/17} \left(\frac{\Sigma(r,t)}{1000{\rm g/cm}^2}\right)^{7/17} {\rm km.}
\end{equation}
The curve for $\tilde{s}_{\rm smax}(r,t)$ at $t =$1Myr is shown on Figure
\ref{fig:smaxnosublim}, and correctly marks the transition 
between the turbulent and gravitational regime at the time 
considered. As time progresses and $\Sigma(r,t)$ decreases so does the 
transition size.

\subsection{Evolution of the solid mass fraction prior to gap opening}
\label{subsec:mathsolids}

The evolution of the solid mass fraction is governed by particle 
diffusion and drift. The analytical prescription used to describe 
the particle size distribution function is particularly useful 
since it can easily be integrated to yield the bulk motion properties, 
as seen in equations (\ref{eq:sceff}) and (\ref{eq:upbar}).
Given the asymptotic behavior of the functions $I(x)$ and $J(x)$, $u_{\rm p}$
is roughly equal to 
\begin{eqnarray}
u_{\rm p} &=& u - 2\eta v_{\rm K} \frac{2\pi \Stmax}{3}  \mbox{   for   } 
\Stmax \ll 1 \mbox{   ,   } \nonumber \\
u_{\rm p} &=& (u - 2\eta v_{\rm K})\frac{\sqrt{\pi}}{4\sqrt{\Stmax}} 
\mbox{   for   } \Stmax \gg 1\mbox{   ,   }
\end{eqnarray}
so that, as expected, the bulk radial velocity of the particles is close to 
that of the gas for small particles, and tends to $0$ when $\smax$ grows. 
Note that $u_p \propto 1/\sqrt{\Stmax}$ instead of $ 1/St(s)$, which accounts 
for the fact that even though particles of size $\smax$ may be largely 
decoupled from the gas, a non-negligible mass fraction is contained in rapidly 
drifting intermediate-size particles. This is the main difference between 
this model and
a single-size particle model; it accounts for the fact
that even when $\Stmax \gg 1$, a significant 
fraction of the collisional encounters are destructive and result in the 
erosion of the larger bodies 
into smaller rapidly drifting particles with $St(s) \simeq 1$. 

\subsubsection{Reservoir of small grains at large radii}
\label{subsubsec:reservoir}

\begin{figure}[ht]
\epsscale{2}
\plotone{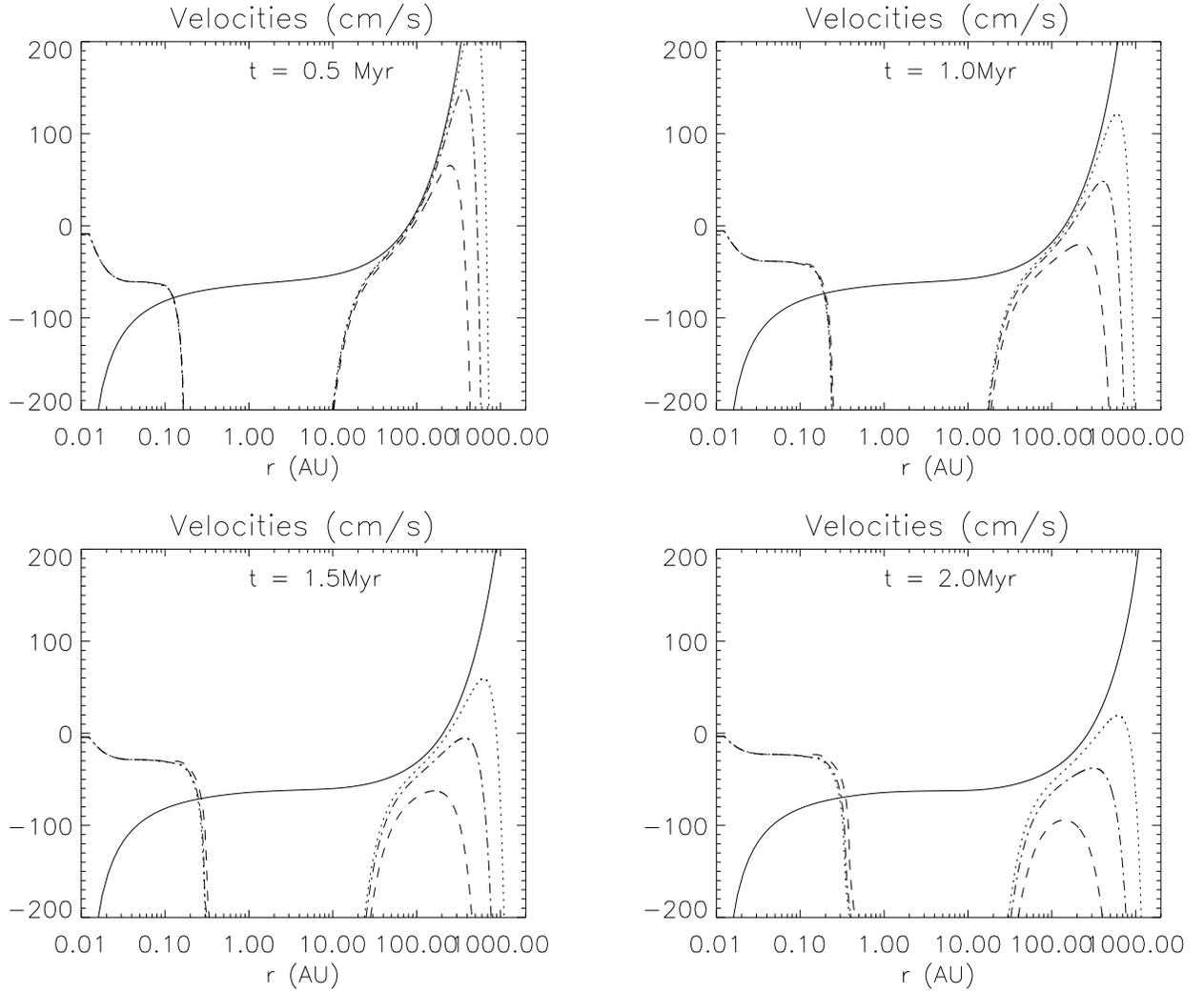}
\caption{Radial velocity $u$ of the gas (solid line) and mass-weighted average
radial velocity $\up$ of particles 
of maximum size $s_{\rm max0}=1\mu$m (dotted line), $3\mu$m (dot-dashed line) 
and $10\mu$m (dashed line). Note how the slow erosion and eventual release 
of the reservoir (i.e. of the spatial region with $\up > 0$) depends 
on $s_{\rm max0}$.}
\label{fig:velocities}
\end{figure}

The sign of $u_{\rm p}$ determines whether grains are transported 
inward or outward.
As expected from equation (\ref{eq:ugasapprox}) the gas velocity changes sign at 
$r_{\rm v}(t)$. This critical radius grows 
linearly with $T$ (defined in equation (\ref{eq:approxsigma})), and therefore 
sweeps outward roughly on the viscous timescale $\tau_{\rm v}$. As a result
for $r < r_{\rm v}$ the particle velocity is necessarily negative, while for 
$r > r_{\rm v}$ particles can be entrained outward provided they are 
strongly coupled with the gas. Since $r_{\rm v}$ is typically much larger 
that $r^{\rm II}_{\rm III}$, all the particles outward of $r_{\rm v}$ are particles 
of size $s_{\rm max0}$. Combining these facts implies that there exists 
a reservoir of small particles at large radii, slowly 
eroded by the outward motion of $r_{\rm v}(t)$. In addition, as the gas 
density drops, the small particles gradually decouple from the gas implying 
that the reservoir begins to ``leak''. Eventually, even the 
smallest particles decouple from the gas, and all the solids come rushing
inward. The evolution of the particle velocity can be seen in Figure 
\ref{fig:velocities}. Note the existence of the reservoir 
(a region of significant radial extent with $\up > 0$) 
for all particle sizes at early time. As time progresses, 
the reservoir begins to ``leak'' as the particles gradually decouple 
from the gas, until a a critical point where $\up$ becomes negative 
for all radii. The phenomenon depends strongly on the 
initial size of the particles $s_{\rm max0}$: three simulations are
shown in which the maximum particle size is respectively $1\mu$m, $3\mu$m and
$10\mu$m. The timescale $t_{\rm p}$ for the release of the particle 
reservoir is clearly much 
shorter for the larger particles (0.84Myr for 10$\mu$m-size particles 
instead of 2.33Myr for $\mu$m-size particles).

\begin{figure}[ht]
\epsscale{1}
\plotone{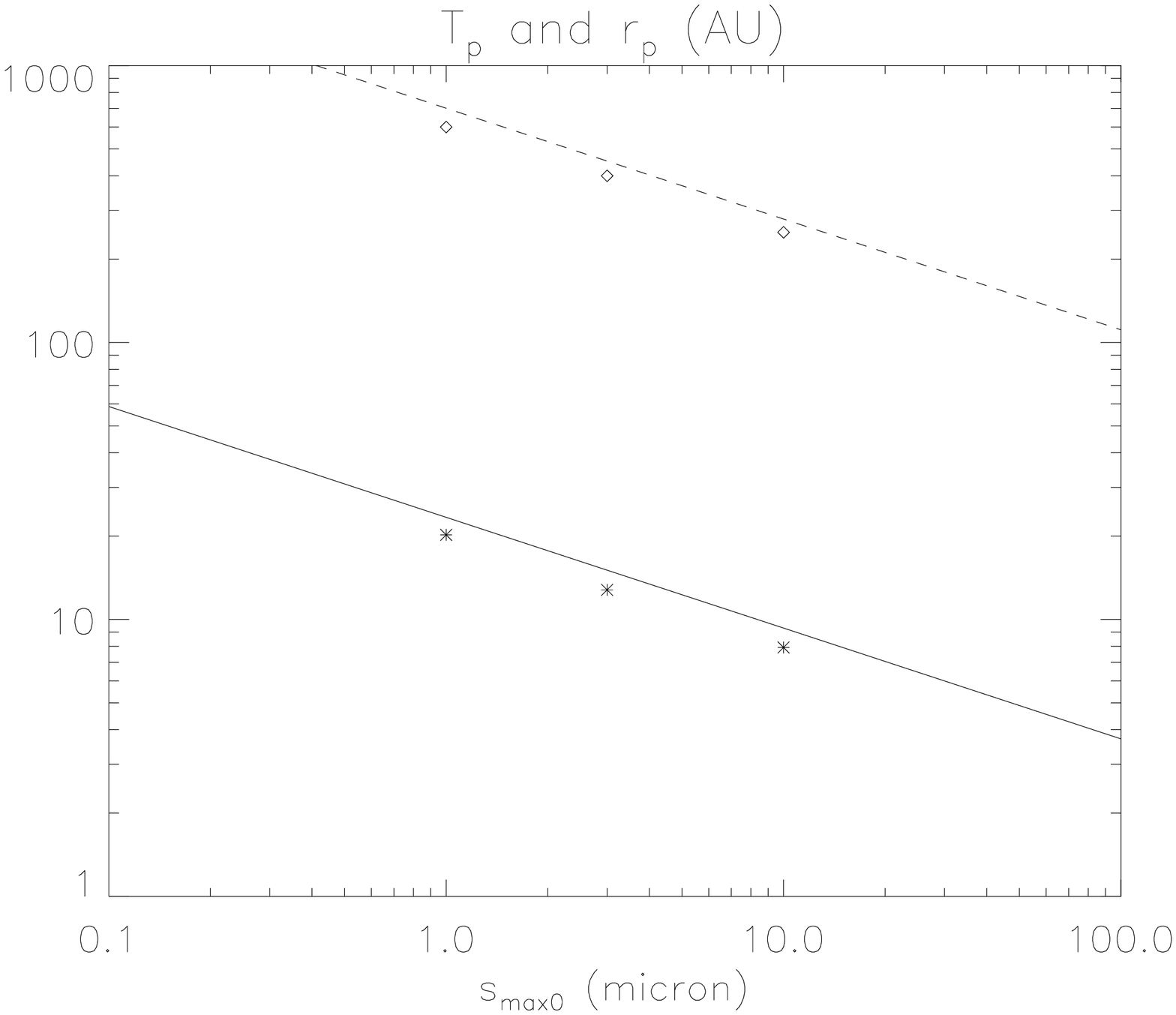}
\caption{Non-dimensional timescale $T_p$ for the release of the particle
 reservoir as a function of initial particle size $s_{\rm max0}$. The 
analytical estimate (solid line) is compared with hand-checked values 
of $T_{\rm p}$ for three numerical simulations of the fiducial model 
with different initial particle sizes (stars). On the same plot is shown the 
numerically determined radius of the reservoir at release $r_p$ in AU (diamonds)
as well as the corresponding analytical estimate $R_0 T_p$ (dashed line).}
\label{fig:Tp}
\end{figure}

In fact, this timescale can easily be estimated by solving 
simultaneously the equations $u_{\rm p} = 0$ and 
$\partial u_{\rm p}/\partial r = 0$. Assuming that 
$\Sigma(r,t)$ is equal to the self-similar solution, 
and that the particles in that regime satisfy 
$\Stmax \ll 1$ (which was checked numerically), 
$u_p \simeq u - 4\pi\eta\Stmax/3$. It follows that  
\begin{equation}
\frac{t_{\rm p}}{\tau_{\rm v}} + 1 = T_{\rm p} = \left[\frac{3}{16 e \pi^2 \eta(R_0)}
\frac{\taud(R_0)}{\tau_{\rm v}} \frac{1}{St_{\rm max0}}\right]^{2/5} \mbox{   ,   }  
\end{equation}
where $\taud(R_0)$ is the dynamical time at $R_0$, $\eta(R_0)$ is
obtained by applying equation (\ref{eq:eta}) at $r=R_0$, $St_{\rm max0}$ is
the Stokes number of particles of size $s_{\rm max0}$ at $R_0$ (see equation
(\ref{eq:stmax0})). Note that in order to derive this expression, I have also made
explicit use of the fact that $q= -1/2$. The quality of this estimate
is found to be excellent given the approximations made (see Figure
\ref{fig:Tp}), and small discrepancies are attributed to the fact that
$\Sigma(r,t)$ is not exactly equal to the self-similar solution, and that 
$\up$ has been approximated by its Taylor expansion for small $\Stmax$. In
the same analytical calculation, the radius of the reservoir at
release is found to be $r_p = R_0 T_p$; checking this solution against
the numerical runs also reveals excellent agreement.

\subsubsection{Evolution of the total mass of the particle disk}
\label{subsubsec:totalmassp}

Since particles in the inner disk can achieve significant sizes, the drift
timescale of grains  within $r_{\rm v}(t)$ is usually 
much smaller than the viscous accretion timescale and/or the age of the disk. 
This can be readily 
seen in Figure \ref{fig:velocities}.

\begin{figure}[ht]
\epsscale{2}
\plottwo{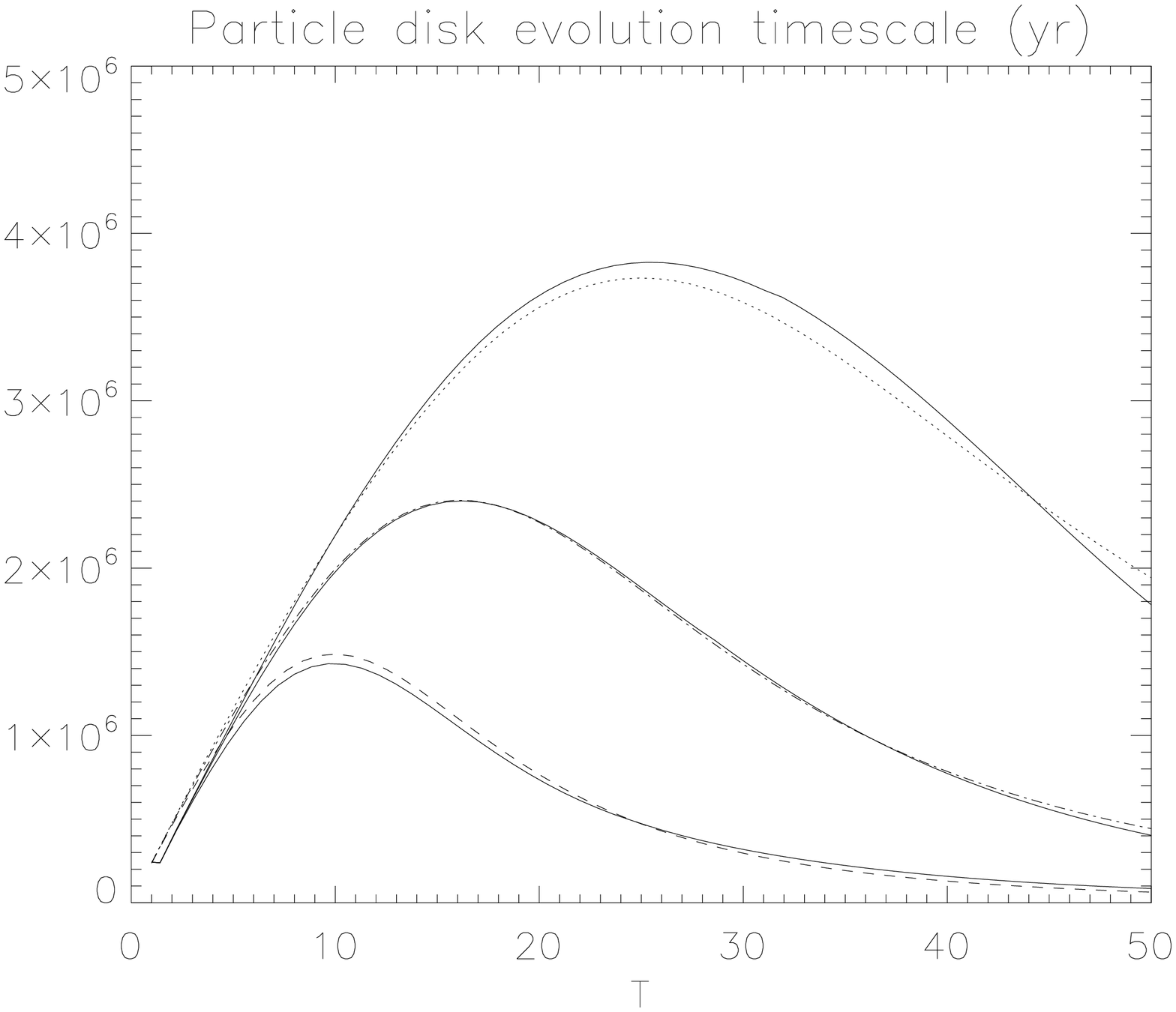}{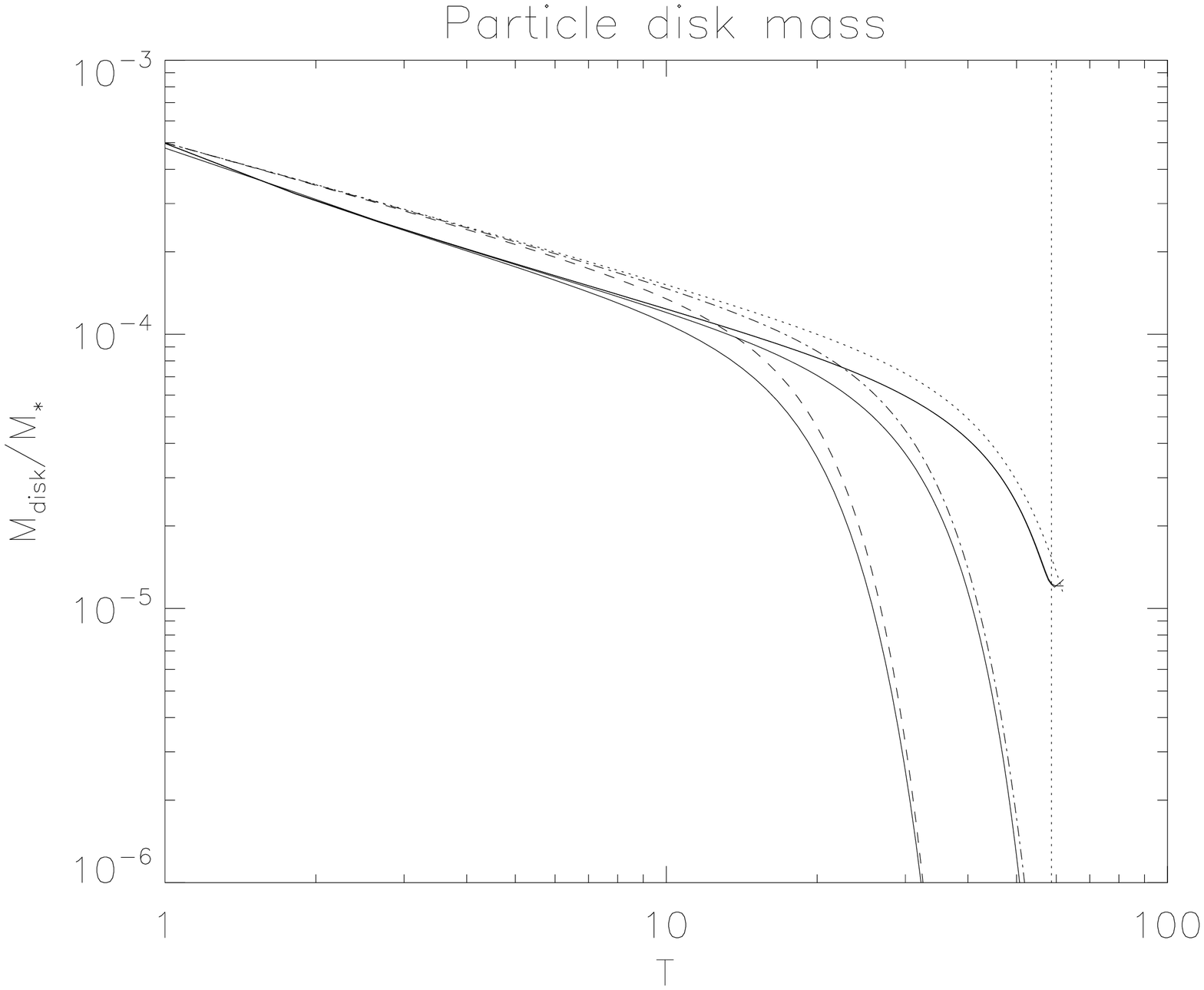}
\caption{Left figure: Particle disk evolution timescale 
$M_{\rm p}/|\dot{M}_{\rm p}|$. 
The solid lines are the outcomes of the numerical simulations for the 
fiducial model with initial particle sizes (from top to bottom) 1$\mu$m, 
3$\mu$m and $10\mu$m respectively. The empirical analytical fits to 
these formula are also shown, with the dotted line for the 1$\mu$m case, 
the dot-dash line for the 3$\mu$m case and the dashed line for the 
10$\mu$m case. Right Figure: Total mass in solids in the disk as a 
function of the nondimensional time $T$. The three solid lines are 
the outcomes of the numerical simulations for the fiducial disk 
model with initial particle sizes (from right to left) 1$\mu$m, 
3$\mu$m and $10\mu$m respectively. Approximates to these numerical 
results are obtained by seeking the solution to equation 
(\ref{eq:dMpdT}), 
and also shown here (linestyles are the same as in the left-side figure).}
\label{fig:taup}
\end{figure}

This very simple fact has important consequences: it implies that the
distribution of solids in the inner disk is uniquely controlled by 
the mass flux leaking from the reservoir. One way to see this
is to look at the particle disk evolution timescale 
$M_{\rm p}/|\dot{M}_{\rm p}|$ obtained by numerical integration of the 
fiducial model. Figure \ref{fig:taup} shows results for the three different 
initial particle sizes considered in the previous section (as solid lines). 
The linear increase for early times ($T \ll T_{\rm p}$) 
mirrors the gas evolution timescale, as expected from the tight 
coupling between the reservoir particles and gas:
\begin{equation}
\dot{M}_{\rm p}(t) = Z_0 \dot{M}(t) = - \frac{Z_0 M_0}{2\tau_{\rm v}} T^{-3/2} \mbox{   ,   }  
\label{eq:mpdot0}
\end{equation}
so that $M_{\rm p}/|\dot{M}_{\rm p}| \simeq M/\dot{M} = 2 \tau_{\rm v} T$. 
As $T$ exceeds $T_{\rm  p}$, the linear increase saturates then rapidly 
turns over, as expected from the release of the reservoir particles.

Constructing an exact analytical model governing the 
evolution of the particle disk after $T_{\rm p}$ from first 
principles turned out to be rather difficult. However, it is possible 
to gain insight into the problem by inspecting the results of the numerical 
simulations first. At later time, one can expect the dynamics of 
the particle disk to 
depend on the reservoir release timescale $T_{\rm p}$. I seek a functional form 
of the kind
\begin{equation}
\frac{M_{\rm p}}{\dot{M}_{\rm p}} \simeq 2 \tau_{\rm v} T f\left(\frac{T}{T_{\rm p}}\right)\mbox{   ,   }  
\end{equation}
with 
\begin{equation}
f(x) = \frac{e^{-a_1 x}}{1 + a_2 x^{a_3}}\mbox{   ,   }  
\end{equation}
(with $a_1$, $a_2$, $a_3 > 0$) which satisfies the requirement that 
$f(x) \rightarrow 1$ as 
$x \rightarrow 0$, and $f(x) \rightarrow 0$ nearly exponentially 
as $x \rightarrow \infty$). A fairly good (but clearly not perfect) 
fit for all three curves is empirically found to have $a_1 = 0.2$, 
$a_2 = 0.25$ and $a_3 = 3.3$, 
implying 
\begin{equation}
\frac{\dd {M}_{\rm p}}{\dd t} \simeq - \frac{M_{\rm p}}{2 T} \exp\left(0.2\frac{T}{T_{\rm P}}\right) \left( 1 + 0.25 \left(\frac{T}{T_{\rm P}}\right)^{3.3}\right)\mbox{   .   }  
\label{eq:dMpdT}
\end{equation}
The fitting curves, for each of the three initial particle sizes chosen, are 
also shown in Figure \ref{fig:taup}a. The initial linear rise as well as 
the maximum are very well represented, while the fit at later times 
(for $T > T_{\rm p}$), in particular for the smaller particle sizes, 
is slightly poorer\footnote{This is again attributed to the numerical boundary 
effects plaguing the gas disk evolution for $t >$ 6Myr, which introduce 
some non-self-similar effects in the solution.}. Integrating equation 
(\ref{eq:dMpdT}) yields an estimate for the total particle disk 
mass as a function of time, which is compared with the results of the 
numerical simulations in Figure \ref{fig:taup}b. The approximate 
solutions follow the trend of the exact solutions well, with some 
small acceptable systematic deviations at early times (see below). 
In particular, it reproduces well the 
very rapid decrease in the particle disk mass as the reservoir is 
finally released ($T > T_{\rm p}$). 

A very important consequence of the ``leaky reservoir'' model is that the total
disk mass is reasonably independent of the physical phenomena taking 
place in the inner disk (provided the bulk drift timescale of the particles
is smaller than the viscous accretion timescale). 
This implies that the evolution of the 
{\it total} disk mass depends very weakly on the particle growth 
rate (and in particular of the sticking efficiency $\epsilon$), and 
of sublimation or condensation fronts. This can actually be seen in Figure 
\ref{fig:taup}, which shows the numerical solution for the fiducial model 
with sublimation/condensation in addition to the three curves discussed 
earlier. The disk mass in the fiducial model is practically indistinguishable 
from that of the disk without sublimation/condensation 
(for the micron-size particles). 

Finally, note that the proposed evolution equation for $M_{\rm p}$ breaks
down if the bulk drift timescale of the particles
is comparable to or larger than the viscous accretion timescale 
or the age of the disk (whichever is smaller).
For instance, if the particles remain small at all times, then they 
naturally follow the evolution of the gas at all times 
(which explains the results of AA07).
As an other example, one can see in Figure \ref{fig:taup} when
$T \rightarrow 1$ that there is a very small difference between the fiducial 
model with sublimation/condensation lines and the model without. This arises 
because at early times, the drift timescale of the particles is much smaller 
than the age of the disk and the solid mass content has not yet had time to 
equilibrate. As a result, most of the solid mass in the inner disk rapidly 
drifts toward the central star. In the absence of 
sublimation lines all of this mass is lost, whereas a significant fraction 
of it can get trapped by the sublimation lines if sublimation/condensation 
is taken into account (see \S\ref{subsec:effsublim}). 

\subsubsection{Evolution of the particle surface density}

Having characterized the global evolution of the total disk mass in solids 
(which was shown to depend only on the initial grain size 
$s_{\rm max0}$ and on 
the viscous diffusion time $\tau_{\rm v}$), 
one could hope to describe the evolution of the surface density of 
grains as well. As mentioned earlier, when the surface density of 
grains within the disk is controlled by the mass flux coming from the 
reservoir one can expect that 
\begin{equation}
2 \pi r u_{\rm p}(r,t) \Sigmap(r,t) \simeq \dot{M}_{\rm p}(t) \mbox{   ,  } 
\label{eq:spapprox}
\end{equation}
where $\dot{M}_{\rm p}(t)$ is given by equation (\ref{eq:dMpdT}).
This approximation turns out to be quite good (except outside 
of $r_{\rm v}(t)$ of course). Unfortunately, the problem lies 
elsewhere: even though $\dot{M}_{\rm p}(t)$ is known, it is particularly  
difficult to estimate $u_{\rm p}(r,t)$ since it 
depends on $\smax(r,t)$ and $\Sigma(r,t)$: the analytical 
estimate of $\smax(r,t)$ given in equation (\ref{eq:approxsmax})  
 is unfortunately not valid in regions I and most of region II 
beyond a few times $10^4$ yr and the self-similar solution 
for $\Sigma(r,t)$ is also invalid for $t > \tau_{\rm v}$ in the same regions.
Using these estimates despite their poor quality yield predictions 
that are off by up to an order of magnitude (see Appendix C). Note that if all 
that is needed is a ``quick and dirty'' order of magnitude estimate 
then the procedure described in the Appendix could be considered satisfactory.
In particular, it does reproduce well the particle surface density dip 
observed in the intermediate disk regions, a feature which could be used 
together with spatially resolved disk observations to constrain 
the value of the sticking efficiency $\epsilon$.

\subsection{Evolution of the solids posterior to gap opening}

The study of the evolution of solids after the gap opening phase
was the main purpose of the work of AA07. 
When growth is ignored, AA07 showed that the evolution of the 
surface density of solids follows that of the gas 
(for small particles).

After the formation 
of the gap, both gas and solids within quickly accrete into
the central star on the clearing timescale, leaving a hole clear 
of both dust and gas. Near the edge of the hole thus formed, an 
inversion in radial pressure gradient causes the particles to drift outward 
instead of inward. As the hole grows all of the grains
outside of $r_{\rm hole}(t)$ are slowly shepherded outward with it.
As a result of these processes,  
most of the solid mass at the time of gap opening 
is retained in the disk but moved to larger and larger radii.

When grain growth is taken into account on the other hand, 
a different picture emerges. The particles within the initial gap radius 
have typically grown to embryo sizes, so that their Stokes number 
is well above unity by the time the gap opens. 
The reduction in the surface density of the gas 
does not affect their drift velocity much (which is exactly the opposite case 
to the AA07 model) and they stay in place while the gas 
within the gap accretes. The ``hole'' thus created still contains a 
significant amount of solid material. 

As direct photo-evaporation takes over, the hole in the gas 
widens as expected but the corresponding reverse radial pressure 
gradient has very little effect on the large particles. 
As a result, one observes
a significant amount of solid material remaining in the inner disk all the way 
out to about 10 AU (in the case of the fiducial model). 
Eventually, the edge of the hole retreats out to regions where the particle 
Stokes number is of order unity, at which point the clearing 
begins as in the AA07 model. As already suggested by AA07, 
all particles smaller than 1-10cm are entrained to large radii, 
while all particles larger than this particular size remain behind.

The 
prediction for the  particle size given in equation (\ref{eq:approxsmax})  
can be used to estimate the radius outside of which particles are entrained,  
by setting the age of the disk to be $t = \tau_{\rm gap}$: let $r_{\rm inner}$ 
the innermost extent of the cleared region, so that
\begin{equation}
r_{\rm inner} = \left[ \left(\frac{\tau_{\rm gap}}{1 {\rm yr}}\right)^2 Z_0^2 
\epsilon^2 \alphat \left(\frac{M_\star}{M_\odot}\right) \right]^{1/3} {\rm AU}
\label{eq:rinner}
\end{equation}
which yields a prediction of $r_{\rm inner} = 16$ AU for the fiducial model. 
This value is larger than that observed in Figure \ref{fig:fiducialsd} by a 
factor of about $2\simeq 10^{1/3}$, which can be attributed to the fact that the estimated 
particle size at this point is a factor of 10 too large compared with the true 
numerical simulations (see Figure \ref{fig:fiducialsmax}). However, 
the estimate in equation (\ref{eq:rinner}) can be thought of as a 
solid upper limit for the radial 
extent of the remaining solid material in the inner regions of the 
disk after the clearing of the gas.

Material outside of $r_{\rm inner}$ is shepherded outward with the 
retreating hole. As more and more material is swept and entrained 
outward, a strongly localized surface density peak appears, which 
eventually grows to be a large as the local surface density of the 
gas. When this happens, two situations could occur within similar 
outcomes: either gravitational instabilities set in, resulting in 
the in-situ formation of a planet, or the frictional drag between 
the particles and the gas begins to influence the evolution of the 
gas itself, and the particles stop moving outward (neither effects 
are included in the numerical model, and therefore cannot be seen 
in the simulations). This effect is difficult to quantify in the 
general case, since neither $\Sigmap$ nor $\Sigma$ are known 
analytically at this stage of the disk evolution. In the fiducial 
model show in Figure \ref{fig:fiducialsd}, this occurs at about 200AU.
This process could lead to the systematic formation of localized 
debris rings reasonably far from the central star (from a few tens 
to a few hundreds of AU), without any need for prior
planet formation or other clumping mechanism. Such debris rings 
are commonly observed, or inferred from the dust dustribution 
(see for instance Schneider {\it et al.} 2006 for direct detections, and
 Strubbe \& Chiang 2006 for inference from spatial dust distribution).

As in the AA07 model, most of the solid content present in the 
disk at $t = \tau_{\rm gap}$ remains in the disk after complete 
clearing of the gas. The fraction of solid material moved to large 
radii depends on the details of the surface density distribution 
of particles at $\tau_{\rm gap}$ (which is not well-known a priori), 
but is typically of the order of 80\%-90\% of the total mass of 
solids; the remaining fraction can be found in the inner disk.

\subsection{Effects of sublimation and condensation}
\label{subsec:effsublim}

The role of sublimation lines on the local accumulation and growth 
of particles has already been shown and discussed by others 
(Stepinski \& Valageas, 1997; Ciesla \& Cuzzi, 2006). Roughly speaking, 
the idea is that particles composed mainly of a given 
chemical species drift inward until they reach the 
sublimation line where they are transformed into vapor. The vapor
diffuses inward much more slowly than the rate of migration of the 
incoming solids, leading to a large enhancement of the local metallicity. 
Through turbulent diffusion, a fraction of the vapor content actually finds 
it way back through the sublimation line and recondenses. The typical 
width of the region where this effect dominates can be evaluated from a local
diffusive lengthscale, and naturally scales as $h(r)$. The exact amount of 
material accumulating in the region is more difficult 
to estimate a priori, since it depends on the difference between the 
solid mass flux 
into the sublimation line and the vapor mass flux out of the sublimation 
line. However, for most of the lifetime of the disk 
the flux of both solid and gaseous material is controlled by the 
reservoir at large radii so that the solid mass flux 
is equal to $Z_0 \dot{M}$, which is also roughly equal to the 
mass flux of vapor out of the sublimation zone. This explains why the
relative strength of the surface density peaks compared with the background
curve fails to grow with time after the initial adjustment period, which 
would necessarily occur otherwise. 

The conclusion is that any local surface density enhancement, 
and associated localized peak in the particle size is
created very early in the lifetime of the disk (this is indeed seen 
in Figure \ref{fig:fiducialsmax}b) -- but it is only later, 
when the gas surface density decays and the metallicity increases, 
that gravitational instabilities in the particle layer could set 
in to trigger the planetary formation process.
This picture, should it be correct, also implies that the relative heights 
of the peaks determines a strict sequence of ``alarm clocks'' on planetary 
formation timescales. 

Note that only three species have been selected in the fiducial model. 
Clearly, there will be as many surface density and particle size peaks 
as the number of separate sublimation temperatures. Also, in this particular 
model the background temperature of the disk is fixed. In reality, the 
disk temperature cools significantly as $\dot{M}$ decreases, resulting in the 
inward migration of the various sublimation lines 
(see Garaud \& Lin 2007, for instance). This will also affect the shape of the 
surface density profile in the inner disk; feedback between the disk 
temperature profile and the evolution of the surface density of grains 
will be the subject of future work.

\section{Discussion}
\label{sec:discuss}

\subsection{Discussion of the particle size distribution function.}

The fundamental assumption underlying this work is that of the maintenance 
of a power-law particle size distribution function at all radii, 
throughout the disk lifetime. The assumption is justified exactly 
only if the collisional timescale 
(note: {\it not} the growth timescale, which is naturally a factor 
of $1/\epsilon$ larger) is shorter than the drift 
timescale for each size-bin. 
Whether this is in fact exactly true is certainly unlikely, 
but neither is it particularly relevant. The correct questions that should be asked
are: (i) {\it how far} from equation (\ref{eq:dnds}) is 
the true size-distribution function in a disk, and (ii) how 
do deviations from (\ref{eq:dnds}) impact the conclusions from this paper?

Question (i) is a fundamental question, with implications 
reaching far beyond the scope of this paper. Attempts at answering it
have come from various angles including both direct observations 
(in disks, but also in molecular clouds, in the ISM, as well 
as in our own solar system) and numerical experiments. As mentioned in 
\S\ref{subsec:sizefunc}, the observational evidence and theoretical 
motivation for a power-law size-distribution function is strong but 
limited to more-or-less spatially isotropic and homogeneous cases where 
there exist no systematic  size-dependent drift or settling velocity 
which could act as a size-filter. Numerical simulations of the 
coagulation-shattering balance in similar conditions also unanimously agree
on the power-law distribution, with indices varying slightly depending 
on the model assumptions but never straying too far from -3.5. 

Unfortunately, there 
has only been one study (Suttner \& Yorke 2001) that self-consistently 
includes a complete parameterization of the coagulation/shattering balance 
together with radial drift and vertical settling of the particles in an 
accretion disk\footnote{The study by Dullemond \& Dominik (2005) 
does not include radial drift, and only treats shattering in a 
very simplistic way.}. Suttner \& Yorke (2001) studied the formation of a 
protostellar accretion disk through the collapse of a uniformly 
rotating molecular cloud core, and closely followed the evolution of
the grain size distribution function at every spatial position 
throughout the collapse phase (first 10,000 years). They found that 
accretion shocks play an important role in limiting the growth of the 
grains; they also found, as expected, that the assumed sticking 
efficiency essentially governs the maximum grain size achievable. 
The dust size distribution functions computed vary strongly with height
above the disk: they show clear evidence that larger grains tend to
be found in the mid-plane, while regions high above the mid-plane 
remain close to the MRN-derived initial conditions. This can be
attributed to a combination of settling of the larger grains as well 
as preferential in-situ growth. The mid-plane regions in their simulations 
appear to be largely depleted of small grains, which would be evidence for
strong deviations from the power-law structure I assume. However, one 
may wonder whether this depletion is indeed true in a real disk, since Suttner
\& Yorke neglect diffusion of the smallest grains by the gas turbulence 
(which could easily bring small grains back towards the mid-plane 
from higher regions of the disk). 

In conclusion, one should bear in mind that 
the assumption made in selecting a power-law size 
distribution function is probably not always strictly justified
(in particular for larger particles). But given the enormous computational 
advantage of this approach, it should be thought of as an acceptable 
trade-off between models in which the full coagulation/shattering equation 
is solved, and models in which only one particle size 
(or a few particle sizes) are considered. 
 
Question (ii) can easily be answered by identifying where in the proposed 
model the assumption of a power-law size-distribution function is used. 

As mentioned in 
\S\ref{subsec:growth}, the minimum particle size $s_{\rm min}$ 
plays {\it no} role in the dynamical evolution of the solids in the disks, 
as long as $s_{\rm min} \ll s_{\rm max}$, which is the case for the MRN 
size-distribution function, and therefore likely to continue being the 
case throughout the disk evolution. Thus whether the smallest grains are 
slightly depleted or not compared with the proposed power-law 
distribution function really does not matter. 

Collisional growth is essentially dominated by encounters between particles 
of similar sizes: even if collisions with smaller particles are more 
frequent, the mass gained is much smaller. This is easily seen 
mathematically in the derivation of the growth rates $\dd \smax/\dd t$ 
in \S\ref{subsec:growth}, where the integral over all possible impactor 
sizes is always dominated by the largest particles, except possibly when 
$\Stmax \gg 1$, in which case the integral is dominated by particles of 
intermediate size. Another way of seeing this is that while the assumptions 
concerning the grain size distribution function used to derive equation 
\ref{eq:groturb} are very different from those of Stepinski \& Valageas (1997) 
who assume that the distribution function is strongly peaked around a single-size,
the outcome is the same whithin some factors of order unity.

The systematic radial motion of decoupled particles ($\Stmax \gg 1$) 
is the only place where 
the assumption made has a significant effect on the model results. If one 
assumes that all particles have a single size $\smax$, then 
$\up(\smax) \propto  1 / \Stmax$ while if one considers the mass-weighted average 
motion of all particles following the proposed size-distribution function 
(\ref{eq:dnds}) then $\up \propto  1 / \sqrt{\Stmax}$ which can 
be significantly larger. As mentioned earlier, this accounts for the fact 
that when a continuum of particle sizes is taken into account, intermediate-size 
grains (with $St(s) \simeq 1)$) do rapidly drift leading to a non-neligible 
mass flux, even if the largest ones are fully decoupled, 

The scaling $\up \propto  1 / \sqrt{\Stmax}$ clearly depends on 
$\dd n/\dd s \propto s^{-3.5}$; whether this exact scaling really applies 
to disks certainly is debatable, but I would argue that the 
general picture of large bodies being eroded by collisions and leading to 
a non-negligible mass-flux even when the larger bodies themselves 
do not drift has to hold. The only caveat is that the 
collision rate becomes null when the bodies reached isolation mass; thus one 
should, for self-consistency, set $\up \rightarrow 0$ in this limit, which 
was not done here (otherwise, smaller particles coming from larger 
radii artificially accumulate in the inner regions). A way forward would be 
to combine the model proposed here with an N-body code, in which 
particles are treated using a size-distribution until they reach embryo 
size, then taken out of the distribution and individually 
followed using N-body simulations. This could be the subject of future work. 


\subsection{Heavy element retention efficiency in the UV-switch model}
\label{subsec:retention}

The simple 
equation for the evolution of the total mass of solids (\ref{eq:dMpdT}) 
can be used to derive the final contents of the disk after complete  
gas dispersal when caused by photo-evaporation from the central star.

If grains in the outer disk remain fully coupled to the gas
 throughout the lifetime 
of the disk (i.e. if $t_p \gg \tau_{\rm gap}$, or equivalently, 
$s_{\rm max0} \ll 1\mu$m) then the amount of material 
left after complete dispersal of the gas can easily be estimated by 
the amount of solids left at $ t = \tau_{\rm gap}$ (see AA07). Therefore
an order of magnitude estimate for the heavy-element retention efficiency is simply
\begin{equation}
\frac{M_{\rm p}(\tau_{\rm gap})}{M_{\rm p0}} \simeq \left( \frac{M_0}{2\tau_{\rm v} \dot{M}_{\rm w} } \right)^{-1/3} \mbox{   ,   }  
\label{eq:Zleft1}
\end{equation}
where $M_{\rm p0} = Z_0 M_0$.
For a fixed initial disk mass, this estimate depends weakly on $R_0$; 
this could explain the very low dispersion observed  in the heavy 
element retention efficiency of evolved systems (Wilden {\it et al.} 2002) 
despite the vastly different disks dispersal timescale required by SED 
observations. 

\begin{figure}[ht]
\epsscale{2}
\plotone{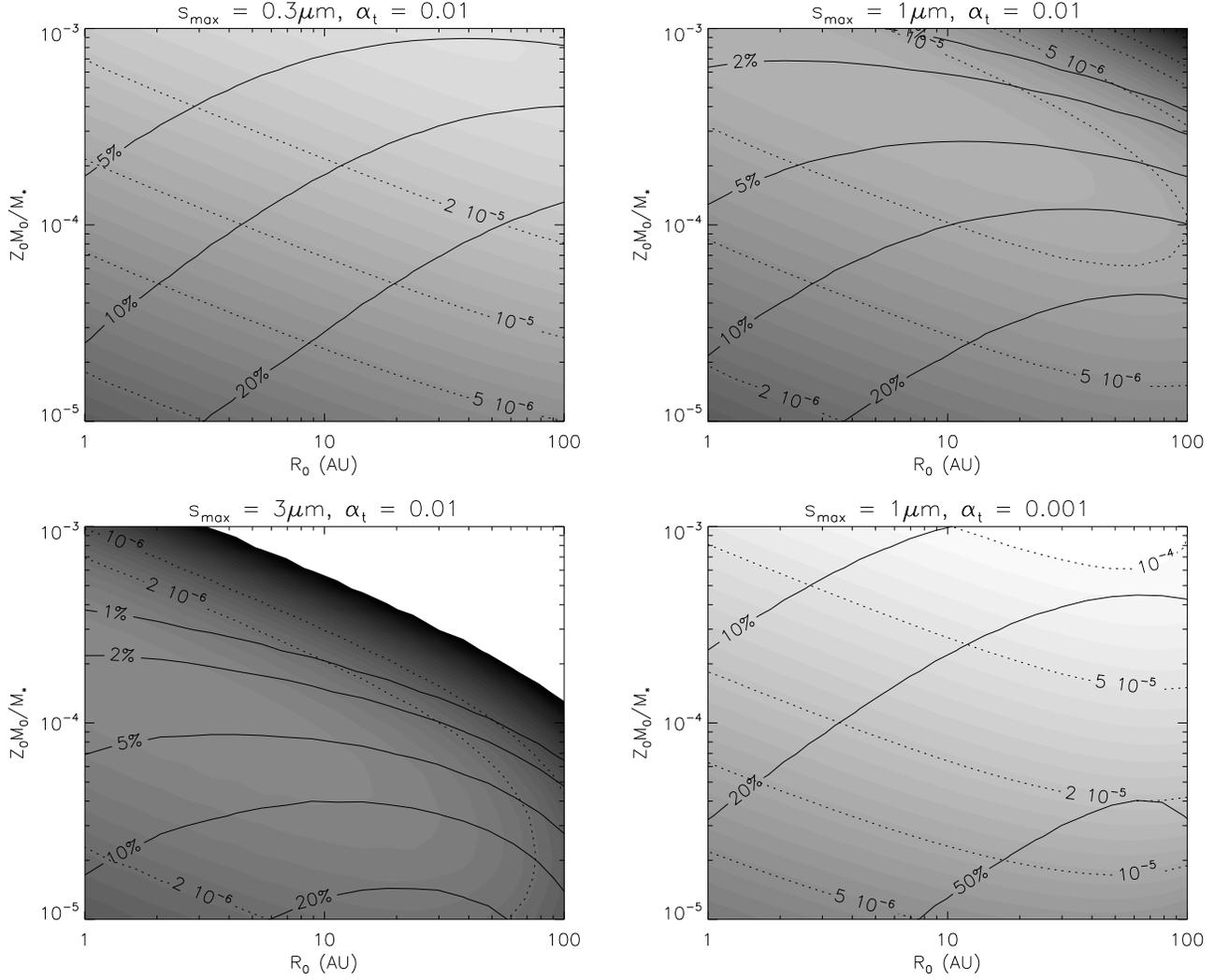}
\caption{Heavy element retention of disks after total photo-evaporation 
of the gas as a function of the initial conditions of the disk, for 
various initial particle sizes and turbulent parameter $\alphat$. The 
color scheme is the same for all four plots. The solid lines 
(at 1, 2, 5, 10, 20 and 50\%) mark the retention efficiency, 
namely the percentage of heavy elements remaining in the disk 
compared with its initial content $Z_0M_0$. The dotted lines 
(at $10^{-6}$, $2\times 10^{-6}$, $5\times 10^{-6}$, $10^{-5}$, 
$2\times 10^{-5}$, $5\times 10^{-5}$, $10^{-4}$ and finally 
$2\times 10^{-4} M_\star$) follow the colored contours and 
mark the actual total mass of the remaining solids. Note 
that $1M_\oplus \simeq 3 \times 10^{-6} M_\odot$ so that 
the $10^{-4}$ contour corresponds to about 33 $M_\oplus$.}
\label{fig:retention}
\end{figure}

To refine this estimate and quantify the effect of the initial grain 
size distribution on the heavy-element
retention efficiency, I integrate (\ref{eq:dMpdT}) from $t = 0$ 
to $t = \tau_{\rm gap}$ for a wide variety of initial conditions 
($M_0$, $R_0$). The results are shown in Figure \ref{fig:retention}. 
Four cases are considered, with varying initial particle size $s_{\rm max0}$ 
and turbulent $\alphat$. The weak dependence on the initial conditions 
of the disk ($M_0$, $R_0$) for given values of $\alphat$ and $s_{\rm max0}$ 
suggested by equation (\ref{eq:Zleft1}) 
is naturally still present: it appears that even with $M_0$ and $R_0$ varying 
by two orders of magnitude, the remaining amount of solids does not 
vary more a factor of a few.  
In the fiducial model for instance ($s_{\rm max0} = 1\mu$m and 
$\alphat = 0.01$) the typical amount of solids left is of the order 
of a few Earth masses for most plausible values of $M_0$ and $R_0$.

As suggested by the results of \S\ref{subsubsec:totalmassp}, the 
retention efficiency drops dramatically for larger initial grain sizes, 
but naturally saturates near the value given by equation (\ref{eq:Zleft1}) 
for very small grain sizes (e.g. compare the results for $s_{\rm max0} = 3\mu$m  with the results for $s_{\rm max0} = 0.3\mu$m). Also suggested by 
(\ref{eq:Zleft1}) is the dependence of the phenomena on $\alphat$: for 
the smaller value of $\alphat = 0.001$, one could expect up to a few 
tens of Earth masses to be left behind in the disk, a value that begins 
to be consistent with the amount of solids left in the Minimum Solar Nebula 
model augmented with the mass of the Oort cloud. Thus it appears that 
consistency of this idea with our own solar system would strongly favour
a model with $s_{\rm max0}$ is no larger than $1\mu$m, and $\alphat$ 
is preferably of the order of 0.001. 

How reliable is this estimate given the simplifications made in the model? 
Conveniently, given that the total disk mass resides mostly in the outer disk, 
and that the amount of material remaining in the inner disk is ultimately 
controlled by the mass flux from the outer reservoir, the mass estimates 
given here are reasonably independent of the physics of the inner disk 
(including sublimation/condensation, but also dead zones, etc..).
There is however an important caveat; as mentioned earlier, large 
protoplanetary embryos which have reached isolation mass
 are fully decoupled from the disk dynamics (both in terms of their
drift velocity and in terms of their collision frequency), and are not 
well-modeled by the size distribution function proposed. The mass contained 
in these embryos could add a few Earth masses to what is presently estimated, 
but only in the inner disk. The total mass of solids which ends up being
shepherded out to the outer solar system is not affected by this problem, 
and is therefore reliably estimated by this method.

\section{Conclusions}
\label{sec:conclude}

This paper presents a new algorithm modeling the evolution of gas and 
solids in protostellar, as well as some reliable quantitative analytical 
estimates for the outcome of the numerical simulations. 

The global disk evolution paradigm is well-reproduced by the numerical 
solutions. Well-known results are recovered, such as the two-timescale 
gas evolution 
(Clarke, Sotomayor \& Gendrin 2001; Alexander, Clarke \& Pringle 2006a), 
the rapid growth of solids in the inner disk (Suttner \& Yorke, 2001, 
Stepinski \& Valageas, 1997, Dullemond \& Dominik, 2005), the shepherding 
of smaller particles by the retreating hole front (Alexander \& Armitage 2007), 
and the accumulation of material near the sublimation lines (Stevenson \& Lunine, 
1988, Stepinski \& Valageas 1997, Ciesla \& Cuzzi, 2006).

Novel conclusions of this paper are:

(i) The evolution of the mass of solids in the disk is essentially controlled
by a reservoir of small grains at large radii. A well-tested empirical formula
for the total solid mass in the disk is given in equation (\ref{eq:dMpdT}). 

(ii) The heavy-element retention efficiency after gas dispersal is controlled 
by the remaining amount of solids left at the time of 
gap opening, and is found to vary weakly with initial disk conditions but 
{\it very sensitively} with initial particle size $s_{\rm max0}$. 
The remaining amount of solids in the Minimum Solar Nebula combined 
with the mass of the Oort cloud is inconsistent with $s_{\rm max0}$ 
greater than 1$\mu$m, and would tend to prefer a value of 
$\alphat \simeq 0.001$. 

(iii) The strong dependence of the gas dispersal timescale on the initial
mass and radius of the disk combined with the weak dependence of the heavy 
element retention efficiency on the same parameters could simultaneously 
explain the wide diversity of the SED observations with the very low
dispersion of the stellar metallicities observed in the Pleiades 
(Wilden {\it et al.} 2002).

(iv) Rapid grain growth in the inner disk implies that solids in the 
form of large planetesimals (with sizes ranging from a few meters to 
1000 km) are locally retained after gas dispersal. The presence of 
a population of large planetesimals in the inner regions of debris disks
has been inferred from interferometric observations of the presence 
of dust despite its very short radiation blowout time (specifically in TW Hya, 
by Eisner, Chiang \& Hillenbrand 2006) 

(v) All the small grains are swept by the retreating gas front at
the edge of the hole and shepeherded outward. When the accumulated surface 
density of grains approaches that of the gas, the gas becomes unable to 
continue moving the grains and will most likely leave them behind. This 
could explain the systematic formation of narrow dust rings at large radii. 

(vi) Regulation of the solid mass flux by the ``leaky reservoir''
implies that any local surface density enhancement (or ``peaks'') 
near sublimation lines 
must be accumulated very early on in the disk lifetime, more precisely
during the initial phase when the disk dynamics are still out of 
equilibrium ($\sim$ first 10,000-100,000 yr). The gradual decay of 
the inner disk gas density through photo-evaporation could then 
trigger gravitational instabilities at well-separated times 
as each peak respectively approaches unit metallicity. 

(vii) Possible constraints on the sticking efficiency of particles 
in the turbulent conditions found in protostellar disks could be
derived from spatially resolved observations of the grain surface 
density of nearby disks (see Appendix C).

Direct comparison of the model predictions with disk SEDs 
is the subject of Paper II. 
\\
\\
{\sc Acknowledgements:} I thank Richard Alexander, Jeff Cuzzi, Tristan Guillot, 
Katherine Kretke, Francis Nimmo, 
Doug Lin and Andrew Youdin for pointing me in the right
direction. I would have liked to acknowledge a funding
source.

\appendix

\section*{Appendix A: Photo-evaporation model} 

Alexander \& Armitage (2007) studied the following model for the 
mass loss rate from photo-evaporation from the central star, based on the works
of Hollenbach {\it et al.} (1994), Font {\it et al.} (2004), as well as 
Alexander, Clarke \& Pringle (2006a and 2006b). 

Prior to the removal of the gas 
in the inner disk, the mass loss rate is caused by the diffuse UV field 
reprocessed high in the disk atmosphere. It is equal to
\begin{equation}
\dot{\Sigma}_{\rm w}(r) = 2 n_0(r) u_{\rm l}(r) \mu m_{\rm H}\mbox{   ,   }  
\end{equation}
where $\mu$ is the mean molecular weight of the ionized gas 
(taken to be $\mu = 1.35$), $m_{\rm H} =1.67 \times 10^{-24}$g 
is the mass of the Hydrogen atom,  the column density at the 
base of the ionized layer $n_0$ is taken to be (see Font {\it et al.} 2004)
\begin{equation}
n_0(r) = 0.14\left(\frac{\Phi}{4\pi \alpha_{\rm B} 
R_{\rm g}^3}\right)^{1/2} \left[ \frac{2}{(r/R_{\rm g})^{15/2} 
+ (r/R_{\rm g})^{25/2}} \right]^{1/5} \mbox{   ,   }  
\end{equation}
with $\Phi$ is the photo-ionizing flux (in photons per second), 
$\alpha_{\rm B} = 2.6 \times 10^{-13}$cm$^3$/s, $R_{\rm g} = G M_*/c_i^2$ 
and $c_i$ is the sound  speed of the ionized gas (taken to be 
10km/s in the fiducial model). The 
launch velocity $u_{\rm l}(r)$ is taken to be (see Font {\it et al.} 2004)
\begin{eqnarray}
u_{\rm l}(r) &=& 0.3423 c_i \exp\left[ -0.3612 \left(\frac{r}{R_{\rm g}} 
- 1\right)\right] \left(\frac{r}{R_{\rm g}} - 1\right)^{0.2457} 
\mbox{   if   } r > 0.1 R_{\rm g} \mbox{   ,   }   \nonumber \\
&=& 0 \mbox{    if   } r < 0.1 R_{\rm g}\mbox{   .   }  
\end{eqnarray}

After clearing of the inner disk, the mass loss rate is mostly caused by 
direct photo-ionization of the inner edge of the remaining disk, 
and accordingly changes to 
\begin{equation}
\dot{\Sigma}_{\rm w}(r,t) = 0.47 \mu m_{\rm H} c_i 
\left[ 1 + \exp\left(-\frac{r - r_{\rm hole}(t)}{h(r_{\rm hole}(t))}
\right)\right]^{-1} \left(\frac{\Phi}{4\pi \alpha_{\rm B} 
r^3_{\rm hole}(t)}\right)^{1/2} \left( \frac{h}{r} \right)^{-1/2} 
\left(\frac{r}{r_{\rm hole}(t)}\right)^{-2.42} \mbox{   ,   }  
\end{equation}
where $r_{\rm hole}(t)$ is the hole radius, well approximated 
to be the radius for which $\Sigma(r,t)$ drops below 
$10^{-7}$g/cm$^2$ (Alexander, private communication).

\section*{Appendix B: Numerical method} 

Equations (\ref{eq:gasevol}), (\ref{eq:vaporevol}) 
(for each species), (\ref{eq:groturb}) or (\ref{eq:grograv}), 
 and finally (\ref{eq:dustevol}) (for each species) 
are evolved simultaneously in time using the following approach.

The independent variables $r$ and $t$ are first normalized 
to 1 AU and to $T_{\rm AU} = 2\pi/\Omega_{\rm K}(1 {\rm AU})$ 
respectively. Following Pringle, Verbunt \& Wade (1986), 
a new variable is introduced
\begin{equation} 
y = r_{\rm AU}^{1/2} \mbox{   ,   }
\end{equation}
upon which space is uniformly discretized. Next, new 
dependent variables $G$, $G^i_{\rm v}$ and $G^i_{\rm p}$ 
are defined as 
\begin{equation}
G = y^{2 q + 4} \Sigma \mbox{  and  } G^i_{\rm v,p} = y^{2 q + 4} 
\Sigma^i_{\rm v,p}\mbox{   .   }
\end{equation}
With these modifications, equations (\ref{eq:gasevol}) and 
(\ref{eq:dustevol}) can be rewritten as 
\begin{equation}
\frac{\partial G}{\partial t_{\rm AU}} = \frac{3 \pi }{2} 
\alphat \sqrt{\gamma}\hbarau^2 y^{2q+1} \frac{\partial^2 G}{\partial y^2} 
- T_{\rm AU} y^{2q+4} \dot{\Sigma}_{\rm w} \mbox{   ,   }
\label{eq:Gevolv}
\end{equation}
(and similarly for each vapor-form species) where $t_{\rm AU} = t/T_{\rm AU}$. 
The evolution equation for the solids becomes
\begin{eqnarray}
\frac{\partial \Gp^i}{\partial t_{\rm AU}} &+& y^{2q+1} \frac{\partial}{\partial y} 
\left[  \left(\frac{\pi}{2} \alphat \sqrt{\gamma}\hbarau^2 (Sc_{\rm eff}^{-1} - 3I) 
\frac{\partial \ln G}{\partial y} - \frac{2 \pi \eta J}{y^{2q+3}} \right) \Gp^i \right] 
\nonumber \\ &=& \frac{\pi}{2} \alphat \sqrt{\gamma}\hbarau^2  y^{2q+1} 
\frac{\partial}{\partial y}\left( \frac{1}{\Sceff} \frac{\partial \Gp^i}{\partial y} \right) \mbox{   ,   }
\label{eq:Gpevolv}
\end{eqnarray}
for each solid-form species. This formulation emphasizes the pure 
diffusive terms, which cause numerical instabilities if not 
treated adequately. For simplicity, $I \equiv I(\sqrt{2\pi \Stmax})$ 
and similarly for $J$.

Equations (\ref{eq:groturb}) or (\ref{eq:grograv}) depending on 
the regime considered involve no spatial derivative, and are therefore 
simply integrated forward in time using a simple explicit second-order
Adams-Bashforth scheme. Equations (\ref{eq:Gpevolv}) and
(\ref{eq:Gevolv}) are integrated forward in time using a centered
implicit scheme for the diffusion terms, an upwind first-order scheme
for the advection terms and a second-order Adams-Bashforth scheme for
all other terms. With this particular semi-implicit scheme, 
it is found that using
about 1000-4000 meshpoints (depending on the radial resolution required)
and a typical timesteps $\Delta t \simeq 0.1-10$yr yields a stable and accurate
solution. 

\section*{Appendix C: A quick-and-dirty estimate for $\Sigmap(r,t)$}

\begin{figure}[ht]
\epsscale{0.5}
\plotone{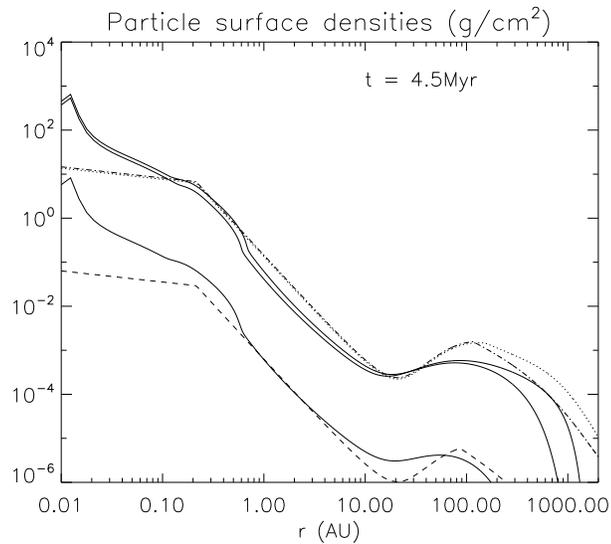}
\caption{True numerical solution (solid lines) and approximate estimates for the total surface density of solids, using the algorithm shown in Appendix C. The estimate for 1$\mu$m size particles is shown as a dotted line, 3$\mu$m as dot-dash line and 10$\mu$m as a dashed line. Note that some of the global trends (in particular the overall normalization of the curve, and the surface density dip) are well-reproduced by the estimate but that quantitative agreement is poor.}
\label{fig:badsigmap}
\end{figure}

To get an estimate for $\Sigmap(r,t)$, one could use
equation (\ref{eq:upbar}) together with equation (\ref{eq:spapprox}). 
Unfortunately, the value of $\Stmax(r,t)$ is unknown since it depends 
both on $\smax(r,t)$ and on $\Sigma(r,t)$, neither of which are particularly 
well approximated in inner and intermediate regions of the disk. Nonetheless, 
let's construct a low-quality estimate of $\Stmax(r,t)$ using the following 
algorithm: at each radius $r$ and time $t$,
\begin{eqnarray}
\smax(r,t) &:=&  \max\left[\frac{\sqrt{2\pi\gamma}}{\rhos} 
\left(\frac{t}{{\rm 1 year}}\right)^2 Z_0^2 \epsilon^2 
\alpha_t \Sigma(r,t) r_{\rm AU}^{-3}  
\left(\frac{M_\star}{M_\odot}\right), s_{\rm max0} \right] \mbox{   ,   } \nonumber \\
m_{\rm iso}(r,t) &:=& \left[2\pi r^2 Z_0 \Sigma(r,t) \tilde{b} \left(\frac{2}{3M_\star}\right)^{1/3} \right]^{3/2} \mbox{   ,   } \nonumber \\
\smax(r,t) &:=& \min\left[\smax(r,t), \left(\frac{3 m_{\rm iso}(r,t)}{4\pi \rhos}\right)^{1/3} \right] \mbox{   ,   } \nonumber \\
\Stmax(r,t)&:=& \frac{\smax(r,t) \rhos}{\sqrt{2\pi\gamma} \Sigma(r,t)}\mbox{   ,   }
\end{eqnarray}
where $m_{\rm iso}(r,t)$ is an estimate of the local isolation mass, and 
where $\Sigma(r,t)$ is given by the self-similar solution (\ref{eq:approxsigma}). 
The resulting predicted solution for $\Sigmap(r,t)$ is shown in Figure 
\ref{fig:badsigmap} at $t = 4.5$Myr, for the three initial particles 
sizes considered in the text (1, 3 and 10$\mu$m). While the global 
features of the solution are fairly well reproduced, the quantitative 
predictions are clearly off by up to an order of magnitude in regions 
I and II of the disk, and even larger in region III where equation 
(\ref{eq:spapprox}) looses validity. 
Note that one of the features that is well-reproduced by the solution 
is the sharp decrease in the surface density of particles in the 
intermediate regions of the disk. This feature corresponds to 
radii where $\Stmax \simeq 1$, which is approximately where
\begin{equation}
r_{\rm dip} = 4.64 \left(\frac{Z_0}{0.01}\right)^{2/3} 
\left(\frac{\epsilon}{0.01}\right)^{2/3} \left(\frac{\alphat}{0.01}\right)^{1/3} 
\left(\frac{M_\star}{M_\odot}\right)^{1/3} \left( \frac{t}{1 {\rm Myr}}\right)^{2/3} 
{\rm AU}\mbox{   .   }
\end{equation}
Interestingly, $r_{\rm dip}$ is entirely independent on the initial 
conditions of the disk ($M_0$ and $R_0$). Since $Z_0$, $M_\star$ and $t$ should be 
fairly well observationally determined by the stellar properties, 
possible measurements of $r_{\rm dip}$ may help constrain the 
product $\epsilon^2 \alphat$ (at least within an order of magnitude).

\end{document}